\documentclass[preprint,prb,showpacs,preprintnumbers,amsmath,amssymb]{revtex4}

\usepackage{graphicx}
\usepackage{dcolumn}
\usepackage{bm}

\newtheorem{thm}{Theorem}[section]
\newtheorem{prop}[thm]{Proposition}
\newtheorem{lem}[thm]{Lemma}

\newtheorem{proposition}[thm]{Proposition}

\newtheorem{rem}[thm]{Remark}
\numberwithin{equation}{section}

\begin{document}

\preprint{APS/123-QED}

\title{On the ground state energy for a magnetic Schr\"odinger operator
and the effect of the De\,Gennes boundary condition}

\author{Ayman Kachmar}
 \altaffiliation[Also at ]{Universit\'e Libanaise, D\'epartement de math\'ematique, Hadeth,
Beyrouth, Liban}
 \email{ayman.kachmar@math.u-psud.fr}
\affiliation{Universit\'e Paris-Sud, D\'epartement de
 math\'ematiques, B\^at. 425, F-91405 Orsay.}



\date{\today}

\begin{abstract}
Motivated by the Ginzburg-Landau theory of superconductivity, we
estimate in the semi-classical limit the ground state energy of a
magnetic Schr\"odinger operator with De\,Gennes boundary condition
and we study the localization of the ground states. We exhibit cases
when  the De\,Gennes boundary condition has strong effects on this
localization.
\end{abstract}

\pacs{Valid PACS appear here}
\keywords{Schr\"odinger operator with magnetic field, semiclassical
analysis, superconductivity}
\maketitle

\section{\label{I}Introduction}
Let $\Omega\subset\mathbb R^2$ be an open bounded domain with
regular boundary. Let us consider a cylindrical superconducting
sample of cross section $\Omega$. The superconducting properties are
described by the minimizers $(\psi,A)$ of the Ginzburg-Landau
functional (cf. Refs.~\onlinecite{deGe, SaSt, Ti})~:
\begin{eqnarray}\label{GL-En}
&& \mathcal G(\psi, A)=\int_\Omega\left\{|(\nabla-i\sigma\kappa
A)\psi|^2 +\sigma^2\kappa^2 |\rm{curl}\,
A-1|^2+\frac{\kappa^2}2(|\psi|^2-1)^2\right\}dx \nonumber\\
&&\hskip1.5cm
+\int_{\partial\Omega}\tilde\gamma\,|\psi(x)|^2\,d\mu_{|_{\partial\Omega}}(x),
\end{eqnarray}
which is defined for pairs $(\psi,A)\in H^1(\Omega;\mathbb C)\times
H^1(\Omega;\mathbb R^2)$. The parameter $\kappa$ is a characteristic
of the material. A material is said to be of type I if $\kappa$ is
sufficiently small and it is said to be of type II when $\kappa$ is
large. The parameter $\sigma$ is the intensity of the applied
magnetic field which is supposed to be constant and perpendicular to
$\Omega$. For a minimizer $(\psi,A)$ of the energy $\mathcal G$, the
function $\psi$ is called the order parameter and $|\psi|^2$
measures the density of superconducting Cooper electron pairs; the
vector field $A$ is called the magnetic potential and ${\rm
curl}\,A$ is the induced magnetic field. Note that the order
parameter $\psi$ satisfies the following boundary condition proposed
by De\,Gennes \cite{deGe}~:
\begin{equation}\label{DeGeBC}
\nu\cdot(\nabla-i\sigma\kappa A)\psi+\tilde\gamma\psi=0,
\end{equation}
where $\nu$ is the unit outward normal of $\partial\Omega$ and
$\tilde\gamma\in\mathbb R$ is called in the physical literature the
De\,Gennes parameter. Note that the boundary condition
(\ref{DeGeBC}) was initially introduced in the theory of PDE by
Robin.\\
The physicist De\,Gennes \cite{deGe} introduced the parameter
$\tilde\gamma$ in order to model interfaces between superconductors
and normal materials. In that context, $\tilde\gamma$ is taken to be
a non-zero positive constant and $\frac1{\tilde\gamma}$ (called the
extrapolation length) usually measures the penetration of the
superconducting Cooper electron pairs in the normal material. The
size of $\tilde\gamma$ depends on the nature of the material
adjacent to the superconductor and it ranges from $\tilde\gamma=0$
(interfaces with insulators) to $\tilde\gamma=+\infty$ (interfaces
with magnetic and ferromagnetic materials). Experiments show that
for superconductors adjacent to ferromagnetic materials, the order
parameter $\psi$ vanishes at the boundary \cite{deGeSa} and the
boundary condition (\ref{DeGeBC}) is changed to the Dirichlet
boundary condition. Negative values of $\tilde\gamma$ were also
considered in the physical literature \cite{FiJo}. It is suggested
that negative values of $\tilde\gamma$ would be useful for modeling
the situation when a superconductor is adjacent to another
superconductor of higher transition temperature.\\
Suppose that we have a type II superconductor (i.e. $\kappa$ is
large). The functional $\mathcal G$ has a critical point of the type
$(0,A)$. Such a critical point is called a normal state. It is then
natural to study whether a normal state is a local minimum of
$\mathcal G$ in the presence of a strong applied magnetic field. The
Hessian of $\mathcal G$ near a normal state is given by~:
$$(\phi,B)\mapsto 2\left[\int_\Omega\left(|(\nabla-i\sigma\kappa
A)\phi|^2-\kappa^2|\phi|^2\right)dx+\int_{\partial\Omega}\tilde\gamma\,|\phi|^2
d\mu_{|_{\partial\Omega}}(x) +(\sigma\kappa)^2\int_\Omega|{\rm
curl}\,B|^2dx\right].$$
By defining the change of
parameter $h=\frac1{\sigma\kappa}$, we have then to study as $h\to0$
the positivity of the quadratic form~:
$$H^1(\Omega)\ni u\mapsto \|(h\nabla-iA)u\|^2_{L^2(\Omega)}
+h^2\int_{\partial\Omega}\tilde\gamma\,|u|^2d\mu_{|_{\partial\Omega}}(x)-(\kappa
h)^2\|u\|^2_{L^2(\Omega)}.$$ The semi-classical limit $h\to0$ is now
equivalent to a large field limit $\sigma\to+\infty$. In order to
study the influence of the size of $\tilde\gamma$, it seems
reasonable to suppose that $\tilde\gamma$ is depending on $h$. Also,
due to the possibility of having different materials exterior to
$\Omega$ together with possible lack of symmetry in the geometry of
$\Omega$, it seems also convenient to take $\tilde\gamma$ as a
function of the boundary.   Thus, given a vector field $A\in
C^\infty(\overline{\Omega};\mathbb R^2)$, a regular real valued
function $\gamma\in C^\infty(\partial\Omega;\mathbb R)$ and a number
$\alpha>0$, let us define the quadratic form~:
\begin{equation}\label{QF}
H^1(\Omega)\ni u\mapsto
q_{h,A,\Omega}^{\alpha,\gamma}(u)=\|(h\nabla-iA)u\|^2_{L^2(\Omega)}+h^{1+\alpha}
\int_{\partial\Omega} \gamma(x)|u(x)|^2d\mu_{|_{\partial\Omega}}(x).
\end{equation}
Observing that $q^{\alpha,\gamma}_{h,A,\Omega}$ is semi-bounded, we
consider the self-adjoint operator associated to
$q^{\alpha,\gamma}_{h,A,\Omega}$ by Friedrich's theorem. This is the
magnetic Schr\"odinger operator $P_{h,A,\Omega}^{\alpha,\gamma}$
with domain $D(P_{h,A,\Omega}^{\alpha,\gamma})$ defined by~:
\begin{equation}\label{OP}
\begin{array}{l}
P_{h,A,\Omega}^{\alpha,\gamma}=-(h\nabla-iA)^2,\\
D(P_{h,A,\Omega}^{\alpha,\gamma})=\{u\in H^2(\Omega);\quad
\nu\cdot(h\nabla-iA)u_{|_{\partial\Omega}}+h^\alpha\gamma
u_{|_{\partial\Omega}}=0\}.
\end{array}
\end{equation}
We denote by $\mu^{(1)}(\alpha,\gamma,h)$ the ground state energy of
$P_{h,A,\Omega}^{\alpha,\gamma}$ which is defined using the min-max
principle by~:
\begin{equation}\label{GSE}
\mu^{(1)}(\alpha,\gamma,h):=\inf_{u\in
H^1(\Omega),u\not=0}\frac{q^{\alpha,\gamma}_{h,A,\Omega}(u)}{\|u\|_{L^2(\Omega)}^2}.
\end{equation}
Let us recall also that this eigenvalue problem is gauge invariant.

In the case when $\gamma\equiv0$ (which corresponds to a
superconductor surrounded by the vacuum), a lot of papers are
devoted to the estimate in a semiclassical regime of the ground
state energy of $P_{h,A,\Omega}^{\alpha,\gamma}$. We would like here
to mention the works of Baumann-Phillips-Tang~\cite{BaPhTa},
Bernoff-Sternberg~\cite{BeSt},
del\,Pino-Felmer-Sternberg~\cite{PiFeSt}, Helffer-Mohamed
\cite{HeMo2}, Helffer-Morame \cite{HeMo3} and the recent work of
Fournais-Helffer \cite{FoHe}. The special case when $\alpha=1$ and
$\gamma$ is a positive constant was considered by Lu-Pan
\cite{LuPa1, LuPa2}. It was shown that in this case the effect of
the De\,Gennes parameter $\gamma$ is weak in the sense that the
limit $\displaystyle\lim_{h\to0}\frac{\mu^{(1)}(1,\gamma,h)}{h}$ is
the same as in the case $\gamma=0$. This regime is therefore not
sufficient to recover all the physically interesting cases
considered in Refs.~\onlinecite{deGeSa, FiJo}. It is the object of
this paper to establish the results announced in
Ref.~\onlinecite{kachN} and to analyze (for all values of $\alpha$)
the influence of the boundary term in (\ref{QF}) on the localization
of the ground state energy of the operator
$P_{h,A,\Omega}^{\alpha,\gamma}$.

 Following the technique of Helffer-Morame~\cite{HeMo3}, we have to
understand the model case of the half-plane when the magnetic field
and the function $\gamma$ are both constant. Consider the magnetic
potential~:
\begin{equation}\label{A0-def}
A_0(x_1,x_2)=\frac12(-x_2,x_1),\quad \forall (x_1,x_2)\in\mathbb
R\times\mathbb R_+.
\end{equation}
Notice that ${\rm curl}\,A_0=1$. Let us define the function
$$\mathbb R\ni\gamma\mapsto\Theta(\gamma),$$
where
\begin{equation}\label{BSp-HP}
\Theta(\gamma):=\inf_{u\in H_{A_0}^1(\mathbb R\times\mathbb
R_+),u\not=0}\frac{\|(\nabla-iA_0)u\|_{L^2(\mathbb R\times\mathbb
R_+)}^2+\gamma\int_{\mathbb R}|u(x_1,0)|^2dx_1}{\|u\|_{L^2(\mathbb
R\times\mathbb R_+)}^2},
\end{equation}
and
\begin{equation}\label{H-A0} H_{A_0}^1(\mathbb R\times\mathbb
R_+) =\{u\in L^2(\mathbb R\times\mathbb R_+);\quad (\nabla-iA_0)u\in
L^2(\mathbb R\times\mathbb R_+)\}.
\end{equation}
Note that $\Theta(\gamma)$ is the bottom of the spectrum of the
operator $P_{h,A_0,\Omega}^{\alpha,\gamma}$ with $h=1$ and
$\Omega=\mathbb R\times\mathbb R_+$. We shall see that
$\Theta(\gamma)<1$ (cf. Theorem~\ref{DaHe}). If $\gamma=0$, we
write~:
\begin{equation}\label{Theta-0}
\Theta_0:=\Theta(0).
\end{equation}
It is  $\Theta_0$ which appears in the analysis for the Neumann
problem \cite{BaPhTa, BeSt, PiFeSt, HeMo2, HeMo3, FoHe, LuPa1,
LuPa2}. Actually, we are interested in the bottom of the spectrum of
the operator $P_{h,A_0,\mathbb R\times\mathbb R_+}^{\alpha,\gamma}$
but a scaling argument gives us~:
\begin{equation}
\label{HP-alhga} \forall h\in\mathbb R_+,\,\forall
\alpha,\gamma\in\mathbb R,\quad \inf{\rm Sp}\left(P_{h,A_0,\mathbb
R\times\mathbb
R_+}^{\alpha,\gamma}\right)=h\Theta(h^{\alpha-1/2}\gamma).
\end{equation}
The semiclassical analysis of the half-plane model depends then on
the sign of both $\alpha-\frac12$ and $\gamma$. We have then to
investigate the asymptotic behavior of $\Theta(\gamma)$ when
$\gamma\to0$ and when $\gamma\to\pm\infty$. This will be the object
of study in Section~\ref{HP}.\\

\noindent Now we state our main results.

\begin{thm}\label{kach1}
Suppose that $\Omega\subset\mathbb R^2$ is open, bounded, connected
and having a smooth boundary. Suppose moreover that the magnetic
field is constant ${\rm curl}\,A=1$. Then, for $\alpha>0$ and
$\gamma\in C^\infty(\partial\Omega;\mathbb R)$,  the ground state
energy of the operator $P_{h,A,\Omega}^{\alpha,\gamma}$ satisfies~:
\begin{equation}\label{kach1th}
\mu^{(1)}(\alpha,\gamma,h)=h\Theta(h^{\alpha-1/2}\gamma_0)\left(1+o(1)\right),\quad
(h\to0),
\end{equation}
where $\gamma_0:=\min_{x\in\partial\Omega}\gamma(x)$.
\end{thm}

Theorem~\ref{kach1} gives a first term approximation of
$\mu^{(1)}(\alpha,\gamma,h)$. The asymptotics (\ref{kach1th}) is
valid without the need to any non-degeneracy hypothesis on the set
of minima of $\gamma$, and holds for the function $\gamma$ being
constant as well. Let us remark that the asymptotics (\ref{kach1th})
depends strongly on $\alpha$. In particular, when $\alpha=\frac12$,
we get~:
$$\lim_{h\to0}\frac{\mu^{(1)}(\alpha,\gamma,h)}{h}=\Theta(\gamma_0)<1,$$
and if $\gamma_0=0$ or if $\alpha>\frac12$, then (cf.
Proposition~\ref{propTay})~:
$$\lim_{h\to0}\frac{\mu^{(1)}(\alpha,\gamma,h)}{h}=\Theta_0<1.$$
When $\alpha<\frac12$, it is the sign of $\gamma_0$ that affects the
asymptotics. Actually, if $\gamma_0<0$ we have (cf.
Proposition~\ref{propAsy}),
$$\lim_{h\to0}\frac{\mu^{(1)}(\alpha,\gamma,h)}{h^{2\alpha}}=-\gamma_0^2,$$
and if $\gamma_0>0$, we have (cf. (\ref{+infty}))~:
$$\lim_{h\to0}\frac{\mu^{(1)}(\alpha,\gamma,h)}{h}=1$$
which is the same behavior as that for the Dirichlet
problem~\cite{HeMo3}. This last regime ($0<\alpha<\frac12$ and
$\gamma_0>0$) is in accordance with the
physical observations in Ref.~\onlinecite{deGeSa}.\\

In the next theorem, we give a two-term asymptotics of
$\mu^{(1)}(\alpha,\gamma,h)$ when $\alpha\in]\frac12,1[$.

\begin{thm}\label{alpha12-1}
Suppose in addition to the hypotheses of Theorem~\ref{kach1} that
$\frac12<\alpha<1$ and that the function $\gamma$ is non-constant.
Then we have the following asymptotic expansion as $h$ tends to
$0$~:
\begin{equation}\label{asy-alp-12-1}
\mu^{(1)}(\alpha,\gamma,h)=h\Theta_0+6M_3\gamma_0h^{\alpha+1/2}
+\mathcal O(h^{\inf(3/2,2\alpha)}),
\end{equation}
where $M_3$ is a strictly positive universal constant.
\end{thm}

The constant $M_3$ satisfies $\Theta'(0)=6M_3$ and it will be
defined precisely in Section~\ref{HP}, see however (\ref{M3}) and
(\ref{der0}). Comparing with the result obtained in
Ref.~\onlinecite{HeMo3}, the second term in the two-term asymptotics
of $\mu^{(1)}(\alpha,\gamma,h)$ when $\gamma=0$ is of order
$h^{3/2}$, whereas it is of order $h^{\alpha+1/2}$ in the regime
considered in Theorem~\ref{alpha12-1}. Let us mention also that in
Ref.~\onlinecite{FoHe}, the authors obtain (when $\gamma=0$) a
complete asymptotic expansion under a generic hypothesis on the
scalar curvature of $\partial\Omega$. It seems that a complete
asymptotic expansion could be obtained in the regime of
Theorem~\ref{alpha12-1} but under the following generic hypothesis
over $\gamma$~:\\
 - $\gamma$ has a finite number of minima;\\
 - all the minima of $\gamma$ are non-degenerate.\\
We leave this point hoping to analyze it in a future work.\\

Next we turn to the question of the localization of the ground
states. Let $u_{\alpha,\gamma,h}$ be a ground state of the operator
$P_{h,A,\Omega}^{\alpha,\gamma}$. We say that $u_{\alpha,\gamma,h}$
is exponentially localized as $h$ tends to $0$ near a closed set
$\mathcal B$ in $\overline\Omega$ if there exists $\beta>0$, and for
each neighborhood $\mathcal V$ of $\mathcal B$, there exist positive
constants $h_0,\delta$ and $C$ such that~:
\begin{equation}\label{defL}
\|u_{\alpha,\gamma,h}\|_{L^2(\Omega\setminus\mathcal V)} \leq
C\exp\left(-\frac{\delta}{h^\beta}\right)\|u_{\alpha,\gamma,h}\|_{L^2(\Omega)},\quad
\forall h\in]0,h_0].
\end{equation}

In the next theorem we describe some  effect of $\gamma$ on the
localization of the ground states of the operator
$P_{h,A,\Omega}^{\alpha,\gamma}$.

\begin{thm}\label{kachLoc}
Under the hypotheses of Theorem \ref{kach1}, if $\gamma_0\leq0$ or
$\frac12\leq\alpha<1$, a ground state  of the operator
$P_{h,A,\Omega}^{\alpha,\gamma}$ is exponentially localized as $h$
tends to $0$
near the boundary points where $\gamma$ is minimum.\\
More precisely, (\ref{defL}) is satisfied with $\beta=1-\alpha$ if
$\gamma_0<0$, $\beta=\frac{1-\alpha}2$ if $\frac12<\alpha<1$, and
$\beta=1/2$ otherwise.
\end{thm}

In the special case $\alpha=1$, the scalar curvature $\kappa_{\rm
r}$ and the function $\gamma$ affects the asymptotic expansion of
the ground state energy to the same order.

\begin{thm}\label{kach-thm-alp=1}
Suppose  in addition to the hypotheses of Theorem~\ref{kach1} that
$\alpha=1$. Then we have the following asymptotic expansion as $h$
tends to $0$~:
\begin{equation}\label{alp=1}
\mu^{(1)}(\alpha,\gamma,h)=h\Theta_0-2M_3(\kappa_{\rm
r}-3\gamma)_{\rm max}h^{3/2}+\mathcal O(h^{13/8}),
\end{equation}
and a ground state $u_{\alpha,\gamma,h}$ of the operator
$P_{h,A,\Omega}^{1,\gamma}$ is localized near the boundary points
where the function
$\kappa_{\rm r}-3\gamma$ is maximal.\\
More precisely, (\ref{defL}) is satisfied with $\beta=1/4$.
\end{thm}

If $\gamma$ is constant, the remainder in (\ref{alp=1}) is better
and of order $\mathcal O(h^{5/3})$. When $\gamma\equiv0$ we recover
in the above theorem the result of Helffer-Morame~\cite{HeMo3}. Let
us mention that the expansion (\ref{alp=1}) is announced by Pan
\cite{Pa03} in the particular case when $\gamma$ is a positive
constant. As in Ref.~\onlinecite{FoHe}, we believe that an
asymptotic expansion with higher terms could be obtained under a
generic
hypothesis on the function $\kappa_{\rm r}-3\gamma$.\\

In the next theorem, we study the case when the function $\gamma$ is
constant and we find that only the scalar curvature plays a role.

\begin{thm}\label{kachLocGC}
Suppose in addition to the hypotheses of Theorem~\ref{kach1} that
the function $\gamma$ is constant and that $\alpha\geq\frac12$.
There exists a constant $M_3(\alpha,\gamma)>0$ such that  we have
the following asymptotic expansion as $h$ tends to $0$~:
\begin{equation}\label{alp>frac12}
\mu^{(1)}(\alpha,\gamma,h)=h\Theta(h^{\alpha-1/2}\gamma)-2M_3(\alpha,\gamma)(\kappa_{\rm
r})_{\rm max}h^{3/2}+o(h^{3/2}).
\end{equation}
Moreover, a ground state  of the operator
$P_{h,A,\Omega}^{\alpha,\gamma}$ is localized as $h$ tends to $0$
near the boundary points where  the scalar curvature is maximal, and
(\ref{defL}) is satisfied with $\beta=1/4$.
\end{thm}

When $\alpha>\frac12$, the constant $M_3(\alpha,\gamma)$ is equal to
the universal constant $M_3$. When $\alpha=\frac12$, we have
$M_3(\frac12,\gamma)=M_3(\gamma)$, where the constant $M_3(\gamma)$
will be defined in Section II (cf. (\ref{M-3-12ga})).

This paper is organized in the following way. In Section~\ref{HP},
we link the analysis of the half-plane model operator to that of a
one dimensional operator. We get in particular the existence of a
number $\xi(\gamma)>0$ such that $\Theta(\gamma)$ is the lowest
eigenvalue of the operator $-\partial_t^2+(t-\xi(\gamma))^2$. Let
$\varphi_\gamma$ be an eigenfunction associated to $\Theta(\gamma)$.
We establish the regularity of $\Theta(\gamma)$ and $\varphi_\gamma$
as functions of $\gamma$, the asymptotic behavior of
$\Theta(\gamma)$ as $\gamma\to\pm\infty$, and uniform estimates with
respect to $\gamma$ describing the exponential decay of
$\varphi_\gamma$ at
infinity.\\
In Section~\ref{Pkach1}, we use the eigenfunction $\varphi_\gamma$
to construct a test function inspired by Refs.~\onlinecite{BeSt,
HeMo3} and we obtain an upper bound for
$\mu^{(1)}(\alpha,\gamma,h)$. We then carry out  a similar analysis
to that in Ref.~\onlinecite{HeMo3} and we use the results of
Section~\ref{HP} to prove Theorem~\ref{kach1}.\\
In Section~\ref{Loc}, we show how to get the localization of the
ground states using Agmon's technique~\cite{Agm}. Finally, in
Section~\ref{curv}, the analysis of a one-dimensional family of
operators on a weighted $L^2$-space appears (cf. (\ref{eq.5.45})).
It is the same family of operators appearing in
Ref.~\onlinecite{HeMo3} (Section~11) but with a different boundary
condition this time. This analysis permits us to derive two-term
asymptotics of the ground state energy showing the influence of the
scalar curvature. We finish then the proofs of
Theorems~\ref{alpha12-1}, \ref{kachLoc}, \ref{kach-thm-alp=1} and
\ref{kachLocGC}.

\section{\label{HP}The model operator}

Given $\gamma\in\mathbb R$, let us consider the quadratic form~:
\begin{equation}\label{QF-HP}
H_{A_0}^1(\mathbb R\times\mathbb R_+)\ni u\mapsto
q[\gamma](u)=\|(\nabla-iA_0)u\|_{L^2(\mathbb R\times\mathbb
R_+)}^2+\gamma\int_{\mathbb R}|u(x_1,0)|^2dx_1.
\end{equation}
The magnetic potential $A_0$ and the form domain $H_{A_0}^1(\mathbb
R\times\mathbb R_+)$ are defined respectively in (\ref{A0-def}) and
(\ref{H-A0}). Observing that  the quadratic form $q[\gamma]$ is
bounded from below, we can associate to $q[\gamma]$, by taking the
Friedrichs extension, a unique self-adjoint operator $P[\gamma]$ on
$L^2(\mathbb R\times\mathbb R_+)$. The min-max principle gives that
the bottom of the spectrum of $P[\gamma]$ is equal to
$\Theta(\gamma)$ (cf. (\ref{BSp-HP})).\\
\subsection{Link with a one dimensional operator}
By a change of gauge and a partial Fourier transformation with
respect to the first variable, we obtain that the spectral analysis
of the operator $P[\gamma]$ will be deduced from that of the
$\xi$-family of one dimensional operators~:
\begin{equation}\label{xiOp}
H[\gamma,\xi]=-\frac{d^2}{dt^2}+(t-\xi)^2,
\end{equation}
with domain
\begin{equation}\label{xiOpDom}
D(H[\gamma,\xi])=\{u\in B^2(\mathbb R_+);\quad u'(0)=\gamma u(0)\},
\end{equation}
where, for a given integer $k$, the space $B^k(\mathbb R_+)$ is
defined by~:
\begin{equation}\label{Bk}
B^k(\mathbb R_+)=\{u\in H^k(\mathbb R_+);\quad t^ku\in L^2(\mathbb
R_+)\}.
\end{equation}
Note that the operator $H[\gamma,\xi]$ has compact resolvent and
hence the spectrum is discrete. We denote by $\mu^{(1)}(\gamma,\xi)$
the first eigenvalue of $H[\gamma,\xi]$. The min-max principle
gives~:
$$\mu^{(1)}(\gamma,\xi)=\inf_{u\in B^1(\mathbb R_+),u\not=0}\frac{q[\gamma,\xi](u)}
{\|u\|^2_{L^2(\mathbb R_+)}},$$ where $q[\gamma,\xi]$ is the
quadratic form associated to $H[\gamma,\xi]$~:
\begin{equation}\label{q[ga,xi]}
q[\gamma,\xi](u)=\int_{\mathbb
R_+}\left(|u'(t)|^2+|(t-\xi)u(t)|^2\right)dt+\gamma|u(0)|^2.
\end{equation}
A spectral analysis using the separation of variables (cf.
Ref.~\onlinecite{ReSiIV}) gives us~:
\begin{equation}\label{F-formula}
\Theta(\gamma)=\inf_{\xi\in\mathbb R}\mu^{(1)}(\gamma,\xi).
\end{equation}
In the following lemma, we collect some useful estimates of
$\mu^{(1)}(\gamma,\xi)$.
\begin{lem}
Given $\epsilon\in]0,1[$, we have,
\begin{equation}\label{Th1.2.2}
\mu^{(1)}(\gamma,\xi)\geq
(1-\epsilon)\mu^{(1)}(0,\xi)-\frac{(\gamma_-)^2}{\epsilon},\quad\forall\gamma,\xi\in\mathbb
R,
\end{equation}
where $\gamma_-=\max(-\gamma,0)$.\\
Moreover, given $\gamma\in\mathbb R$, we have~:
\begin{equation}\label{limxiinfty}
\lim_{\xi\to-\infty}\mu^{(1)}(\gamma,\xi)=+\infty,\quad
\lim_{\xi\to+\infty}\mu^{(1)}(\gamma,\xi)=1.
\end{equation}
\end{lem}
\noindent{\bf Proof.} Using the density of
$C_0^\infty(\overline{\mathbb R_+})$ in $H^1(\mathbb R_+)$, we get
for any $u\in H^1(\mathbb R_+)$~:
\begin{equation}\label{Den-C0H1}
|u(0)|^2=-2\int_0^\infty u(t)u'(t)dt.
\end{equation}
By the Cauchy-Schwarz inequality, we get for any $\alpha>0$~:
$$|u(0)|^2\leq
\alpha\|u\|^2_{L^2(\mathbb R_+)}+\frac1\alpha\|u'\|^2_{L^2(\mathbb
R_+)}.$$ Taking $\alpha=\frac{\epsilon}{\gamma}$ (with $\gamma<0$),
we get~:
\begin{equation}\label{Th1.2.2-qf}
q[\gamma,\xi](u)\geq(1-\epsilon)q[0,\xi](u)
-\frac{(\gamma_-)^2}\epsilon\|u\|^2_{L^2(\mathbb R_+)},\quad \forall
u\in B^1(\mathbb R_+).
\end{equation} The min-max principle now gives (\ref{Th1.2.2}).\\
Notice that (\ref{limxiinfty}) is valid for $\gamma=0$
(Ref.~\onlinecite{HeMo3}). So the limit as $\xi\to-\infty$ in
(\ref{limxiinfty}) is now a consequence of the estimate
(\ref{Th1.2.2}). For the reader's convenience, let us give for
non-zero $\gamma$ a proof for the limit as $\xi\to+\infty$ in
(\ref{limxiinfty}). Let us denote by $\mu^D(\xi)$ the first
eigenvalue of the Dirichlet realization of the harmonic oscillator
$-\partial_t^2+(t-\xi)^2$ on $\mathbb R_+$. We have by the min-max
principle~:
\begin{equation}\label{ga=0-n0}
\mu^{(1)}(0,\xi)+\gamma|\varphi_{\gamma,\xi}(0)|^2\leq\mu^{(1)}(\gamma,\xi)\leq\mu^D(\xi),
\end{equation}
where $\varphi_{\gamma,\xi}$ is the $L^2$-normalized eigenfunction
associated to $\mu^{(1)}(\gamma,\xi)$. Let us notice also
that~\cite{HeMo3, ReSiIV}
$\displaystyle\lim_{\xi\to+\infty}\mu^D(\xi)=1$. So, if we know that
$\displaystyle\lim_{\xi\to+\infty}|\varphi_{\gamma,\xi}(0)|^2=0$,
then (\ref{ga=0-n0}) is sufficient to deduce the limit as
$\xi\to+\infty$ in (\ref{limxiinfty}). Thus, it remains for us to
prove the following claim~:
\begin{equation}\label{Dec-xi-+inf}
\begin{array}{l}
\text{Given }\epsilon\in]0,1[\text{ and }\gamma\in\mathbb R,\text{
there exists a constant } C>0 \text{ such that,}\\
\forall\xi\in[C,+\infty[,\quad |\varphi_{\gamma,\xi}(0)|^2\leq
Ce^{-\epsilon\xi/2}.
\end{array}
\end{equation}
Let us mention that the decay in (\ref{Dec-xi-+inf}) is not
optimal~\cite{BoHe, kach}. We prove (\ref{Dec-xi-+inf}) using Agmon
type estimates~\cite{Agm}. Let $\Phi$ be a regular function with
compact support. An integration by parts gives the following
identity~:
\begin{equation}\label{1eqnAg1-Th}
q[\gamma,\xi]\left(e^\Phi\varphi_{\gamma,\xi}\right)=
\mu^{(1)}(\gamma,\xi)\left\|e^\Phi\varphi_{\gamma,\xi}\right\|^2_{L^2(\mathbb
R_+)}+\left\|\Phi'e^\Phi\varphi_{\gamma,\xi}\right\|^2_{L^2(\mathbb
R_+)}.
\end{equation}
Using the estimate (\ref{Th1.2.2-qf}) (with $\epsilon=1/2$) together
with the fact that $\mu^D(\xi)$ is bounded for $\xi\in\mathbb R_+$,
we can rewrite (\ref{1eqnAg1-Th}) in the form~:
\begin{equation}\label{1eqnAg2-Th}
\frac12\left(\left\|\left(e^\Phi\varphi_{\gamma,\xi}\right)'\right\|^2_{L^2(\mathbb
R_+)}+\left\|(t-\xi)e^\Phi\varphi_{\gamma,\xi}\right\|^2_{L^2(\mathbb
R_+)}\right)\leq \tilde
C\left\|e^\Phi\varphi_{\gamma,\xi}\right\|^2_{L^2(\mathbb
R_+)}+\left\|\Phi'e^\Phi\varphi_{\gamma,\xi}\right\|^2_{L^2(\mathbb
R_+)},
\end{equation}
for some constant $\tilde C>0$. We choose now $\Phi$ as~:
$$\Phi(t):=\left\{\begin{array}{cll}
\epsilon\xi\frac{(t-1)^2}2&if& 0\leq t\leq 1,\\
0&if& t\geq 1.\end{array}\right.$$ Under this choice of $\Phi$, we
can get  a sufficiently large constant $C>0$ such that, for
$\xi\in[C,+\infty[$, we can rewrite (\ref{1eqnAg2-Th}) in the form~:
$$\left\|e^\Phi\varphi_{\gamma,\xi}\right\|_{H^1(\Omega)}\leq C.$$
Using the Sobolev imbedding $H^1(\mathbb R_+)\hookrightarrow
L^\infty(\mathbb R_+)$, this last estimate is sufficient to deduce
(\ref{Dec-xi-+inf}).\hfill$\Box$\\

 Following the analysis of Dauge-Helffer
\cite{DaHe}, we have now the following result.
\begin{thm}\label{DaHe}
For each $\gamma\in\mathbb R$, $\Theta(\gamma)<1$ and the function
$\mathbb R\ni\xi\mapsto\mu^{(1)}(\gamma,\xi)$ attains its minimum at
a unique  positive point $\xi(\gamma)$ that satisfies~:
\begin{equation}\label{relXiGa}
\xi(\gamma)^2=\Theta(\gamma)+\gamma^2.
\end{equation}
\end{thm}
\noindent{\bf Proof.} Let us notice that by Kato's theory
(Ref.~\onlinecite{Ka}), the maps
$$\xi\mapsto\mu^{(1)}(\gamma,\xi),\quad
\xi\mapsto\varphi_{\gamma,\xi}\in L^2(\mathbb R_+)$$ are analytic.
Here we recall that $\varphi_{\gamma,\xi}$ is the unique strictly
positive and $L^2$-normalized eigenfunction associated to
$\mu^{(1)}(\gamma,\xi)$. Let us consider $\tau>0$. Note that,
$$\mu^{(1)}(\gamma,\xi+\tau)\varphi_{\gamma,\xi+\tau}(t+\tau)
=H[\gamma,\xi]\left(\varphi_{\gamma,\xi+\tau}(t+\tau)\right),\quad
\forall t\in\mathbb R_+.$$ Taking the scalar product with
$\varphi_{\gamma,\xi}$ and then integrating by parts, we get~:
\begin{eqnarray}
&&\left(\mu^{(1)}(\gamma,\xi+\tau)-\mu^{(1)}(\gamma,\xi)\right)
\int_{\mathbb R_+}
\varphi_{\gamma,\xi+\tau}(t+\tau)\varphi_{\gamma,\xi}(t)\,dt\label{DaHeeq1}\\
&&=\varphi_{\gamma,\xi+\tau}'(\tau)\varphi_{\gamma,\xi}(0)
-\gamma\varphi_{\gamma,\xi+\tau}(\tau)\varphi_{\gamma,\xi}(0).\nonumber
\end{eqnarray}
Recall that we have the boundary conditions~:
$$\varphi_{\gamma,\xi+\tau}'(0)=\gamma\varphi_{\gamma,\xi+\tau}(0),\quad
\varphi_{\gamma,\xi}'(0)=\gamma\varphi_{\gamma,\xi}(0).$$ Then we
can rewrite (\ref{DaHeeq1}) as~:
\begin{eqnarray*}
&&\frac{\mu^{(1)}(\gamma,\xi+\tau)-\mu^{(1)}(\gamma,\xi)}\tau
\int_{\mathbb R_+}
\varphi_{\gamma,\xi+\tau}(t+\tau)\varphi_{\gamma,\xi}(t)\,dt\\
&&=\left[\frac{\varphi_{\gamma,\xi+\tau}'(\tau)-\varphi'_{\gamma,\xi+\tau}(0)}\tau-\gamma
\frac{\varphi_{\gamma,\xi+\tau}(\tau)-\varphi_{\gamma,\xi+\tau}(0)}\tau\right]\cdot
\varphi_{\gamma,\xi}(0).
\end{eqnarray*}
By taking the limit as $\tau\to0$, we get~:
$$\partial_{\xi}\mu^{(1)}(\gamma,\xi)=
\left(\varphi_{\gamma,\xi}''(0)-\gamma\varphi_{\gamma,\xi}'(0)\right)
\varphi_{\gamma,\xi}(0).$$ Finally, we make the substitutions~:
$$\varphi_{\gamma,\xi}''(0)
=\left(\xi^2-\mu^{(1)}(\gamma,\xi)\right)\varphi_{\gamma,\xi}(0),\quad
\varphi_{\gamma,\xi}'(0)=\gamma\varphi_{\gamma,\xi}(0),$$ and we get
the following formula,
\begin{equation}\label{F-formula'}
\partial_{\xi}\mu^{(1)}(\gamma,\xi)=
\left(\xi^2-\mu^{(1)}(\gamma,\xi)
-\gamma^2\right)|\varphi_{\gamma,\xi}(0)|^2,
\end{equation}
called usually the $F$-formula (cf. Refs.~\onlinecite{DaHe, LuPa3}).
Using (\ref{Th1.2.2}) and (\ref{limxiinfty}), we get~:
$$\partial_\xi\mu^{(1)}(\gamma,\xi)_{|_{\xi=0}}<0,\quad
\partial_\xi\mu^{(1)}(\gamma,\xi)_{|_\xi=\eta}>0,$$
for a sufficiently large $\eta>0$. This gives the existence of a
positive critical point of $\mu^{(1)}(\gamma,\xi)$. Let us notice
now that for any critical point $\xi_c$ of $\mu^{(1)}(\gamma,\xi)$,
we have~:
$$\partial^2_\xi\mu^{(1)}(\gamma,\xi)_{|_{\xi=\xi_c}}=2\xi_c|\varphi_{\gamma,\xi}(0)|^2.$$
This shows that any negative critical point is a global maximum and
any positive critical point is a global minimum of
$\mu^{(1)}(\gamma,\xi)$. Coming back to (\ref{limxiinfty}),
$\displaystyle\lim_{\xi\to-\infty}\mu^{(1)}(\gamma,\xi)=+\infty$,
and thus there does not exist any negative critical points.
Therefore, the minimum of $\xi\mapsto\mu^{(1)}(\gamma,\xi)$ is
attained at a unique  point $\xi(\gamma)>0$ and the function
$\xi\mapsto\mu^{(1)}(\gamma,\xi)$ is strictly increasing on
$[\xi(\gamma),+\infty[$. This proves in particular (recalling
(\ref{F-formula}))~:
$$\Theta(\gamma)=\mu^{(1)}(\gamma,\xi(\gamma))<1.$$
\hfill$\Box$\\

In the sequel, we denote by $\varphi_\gamma$ the unique strictly
positive and $L^2$-normalized eigenfunction associated to the
eigenvalue $\Theta(\gamma)$, and by $H[\gamma]$ the operator
$H[\gamma,\xi(\gamma)]$~:
\begin{equation}\label{phi-ga,H-ga}
\varphi_\gamma=\varphi_{\gamma,\xi(\gamma)},\quad
H[\gamma]=H[\gamma,\xi(\gamma)].
\end{equation}

In the next lemma, we collect various useful relations satisfied by
the eigenfunction $\varphi_\gamma$. These relations are similar to
those given in  Appendix A of Ref.~\onlinecite{HeMo3}.
\begin{lem}\label{Mn-gam}
For each $\gamma\in\mathbb R$, the following relations hold~:
\begin{equation}\label{M1-gam}
\int_{\mathbb R_+}(t-\xi(\gamma))|\varphi_\gamma(t)|^2dt=0,
\end{equation}
\begin{equation}\label{M2-gam}
\int_{\mathbb
R_+}(t-\xi(\gamma))^2|\varphi_\gamma(t)|^2dt=\frac{\Theta(\gamma)}2-\frac\gamma4|\varphi_\gamma(0)|^2,
\end{equation}
\begin{equation}\label{M3-gam}
\int_{\mathbb R_+}(t-\xi(\gamma))^3|\varphi_\gamma(t)|^2dt
=\frac16\left[1-2(\gamma\xi(\gamma))^2\right]|\varphi_\gamma(0)|^2.
\end{equation}
\end{lem}
\noindent{\bf Proof.} We follow the calculations done in
Bernoff-Sternberg \cite{BeSt}. Let us consider the differential
operator~:
$$L=-\partial_t^2+(t-\xi(\gamma))^2-\Theta(\gamma).$$
Note that for any polynomial $p$, we have the following identity~:
\begin{equation}\label{eq1.2.31}
  L(2p\varphi_\gamma-p'\varphi_\gamma)=\left(p^{(3)}-4\left[(t-\xi(\gamma)-\Theta(\gamma)\right]p'
-4(t-\xi(\gamma))p\right)\varphi_\gamma.
\end{equation}
Let $v=2p\varphi_\gamma'-p'\varphi_\gamma$. Integrating by parts we
obtain~:
\begin{equation}\label{eq1.2.32}
\int_0^{+\infty}\varphi_\gamma(t) (Lv)(t)\,dt=(v'(0)-\gamma
v(0))\varphi_\gamma(0).
\end{equation}
Taking $p=1$, we get~:
$$-4\int_0^{+\infty}(t-\xi(\gamma))|\varphi_\gamma(t)|^2dt=2\left(\xi(\gamma)^2-\gamma^2-\Theta(\gamma)\right)|\varphi_\gamma(0)|^2.$$
Recalling (\ref{relXiGa}), the above formula proves (\ref{M1-gam}).\\
We prove (\ref{M2-gam}) by taking $p=(t-\xi(\gamma))$. To prove
(\ref{M3-gam}), we take $p=(t-\xi(\gamma))^2$. Note that we have in
this case~:
$$v'(0)-\gamma v(0)=2\left(2(\gamma\xi(\gamma))^2-1\right)\varphi_\gamma(0).$$
We get now from (\ref{eq1.2.31}) and (\ref{eq1.2.32})~:
$$-12\int_0^{+\infty}(t-\xi(\gamma))^3|\varphi_\gamma(t)|^2dt=2\left(2(\gamma\xi(\gamma))^2-1\right)|\varphi_\gamma(0)|^2.$$
This proves (\ref{M3-gam}).\hfill$\Box$\\

For $\gamma\in\mathbb R$, let us define the parameter~:
\begin{equation}\label{M-3-12ga}
M_3(\gamma)=\frac16\left(1+(\gamma\xi(\gamma))^2\right)|\varphi_\gamma(0)|^2,
\end{equation}
and when $\gamma=0$, we write $M_3:=M_3(0)$. Note that
(\ref{M3-gam}) gives~:
\begin{equation}\label{M3}
M_3=\int_{\mathbb R_+}(t-\xi_0)^3|\varphi_0(t)|^2dt,
\end{equation}
where $\xi_0:=\xi(0)$. The constant $M_3$ is the universal constant
appearing in Theorems~\ref{alpha12-1}, \ref{kach-thm-alp=1}, and the
parameter $M_3(\gamma)$ appears as
$M_3(\frac12,\gamma)$ in Theorem~\ref{kachLocGC}.\\

\subsection{Regularity}
We discuss now the regularity of the functions
$\gamma\to\Theta(\gamma)\in\mathbb R$ and
$\gamma\mapsto\varphi_\gamma\in L^2(\mathbb R_+)$. It seems for us
that  Kato's theory (cf. Ref.~\onlinecite{Ka}) do not apply in this
context at least for the reason that we do not know a priori whether
the expression of the operator
$$H[\gamma]=-\frac{d^2}{dt^2}+(t-\xi(\gamma))^2$$
depends analytically on $\gamma$. Inspired by Bonnaillie~\cite{Bon},
we use a modification of Grushin's method \cite{Gru} and we get the
following proposition.
\begin{prop}\label{reg}
The functions $\mathbb R\ni \gamma\mapsto\Theta(\gamma)\in\mathbb R$
and $\mathbb R\ni\gamma\mapsto\varphi_\gamma\in L^2(\mathbb R_+)$
are
$C^\infty$.\\
Moreover, the function $\mathbb R\ni\gamma\mapsto\varphi_\gamma\in
L^\infty(\mathbb R_+)$ is locally Lipschitz.
\end{prop}
The specific difficulty in proving Proposition~\ref{reg} comes from
the fact that both the expression and the domain of the operator
$H[\gamma]$ depend on $\gamma$. To work with an operator with a
fixed domain, we consider a cut-off $\chi$ that is equal to $1$ on
$[0,1]$ and we apply the invertible transformation $\varphi\mapsto
\tilde\varphi=e^{-\gamma t\chi(t)}\varphi$ that transforms the
boundary condition $\varphi'(0)=\gamma\varphi(0)$ to the usual
Neumann boundary condition $\tilde\varphi'(0)=0$ and
leaves the spectrum invariant (cf. Proof of Proposition~\ref{1Bonn}).\\

In the next proposition we determine $\Theta'(\gamma)$. This is  a
first step in the proof of Proposition~\ref{reg}.
\begin{prop}\label{propTay}
The function $\gamma\mapsto\Theta(\gamma)$ is of class $C^1$ and
satisfies~:
\begin{equation}\label{derG}
\Theta'(\gamma)=|\varphi_\gamma(0)|^2.
\end{equation}
In particular, we have~:
\begin{equation}\label{der0}
\Theta'(0)=6M_3.
\end{equation}
\end{prop}
\begin{rem}\label{derThga}
Using Formula (\ref{relXiGa}) we get also that the function
$\gamma\mapsto\xi(\gamma)$ is of class $C^1$.
\end{rem}
\noindent{{\bf Proof of Proposition~\ref{propTay}.}} Let $\tau$ be a
real number. We shall define the following trial function~:
$$u=e^{\tau t}(\varphi_\gamma+\tau u_1),$$
where
$$u_1=\left(H[\gamma]-\Theta(\gamma)\right)^{-1}
\left\{|\varphi_\gamma(0)|^2\varphi_\gamma+2\varphi_\gamma'
+2(\xi(\gamma+\tau
)-\xi(\gamma))(t-\xi(\gamma))\varphi_\gamma\right\}.$$ By standard
Fredholm theory, the operator $(H[\gamma]-\Theta(\gamma))^{-1}$ is
defined on the orthogonal space of $\varphi_\gamma$ and has values
in $D(H[\gamma])$. Hence, the function $u_1$ is well defined, thanks
to (\ref{M1-gam}), and the function $u$ satisfies the boundary
condition $u'(0)=(\gamma+\tau) u(0)$. When $\tau$ is sufficiently
small, it is a result of the exponential decay of $\varphi_\gamma$
at $+\infty$ (cf. Propositions~\ref{propDecEF} and \ref{propDecReg})
and standard elliptic estimates that $u\in B^2(\mathbb R_+)$.
Therefore, $u\in D(H[\gamma+\tau])$, and we have~:
\begin{equation}\label{Der0-1}
H[\gamma+\tau](u)=e^{\tau t}
\left(-\partial_t^2+(t-\xi(\gamma+\tau))^2-2\tau\partial_t-\tau^2\right)(\varphi_\gamma+\tau
u_1).
\end{equation}
Using the decomposition~:
$$H[\gamma+\tau]=H[\gamma]-2(\xi(\gamma+\tau)-\xi(\gamma))(t-\xi(\gamma))+(\xi(\gamma+\tau)-\xi(\gamma))^2,$$
we can rewrite (\ref{Der0-1}) as~:
\begin{eqnarray}\label{Der0-2}
&&\left(H[\gamma+\tau]-\Theta(\gamma)-|\varphi_\gamma(0)|^2\tau\right)(u)\nonumber\\
&&=\tau^2 e^{\tau
t}\left(\Theta(\gamma)+|\varphi_\gamma(0)|^2)u_1+(\xi(\gamma+\tau)-\xi(\gamma)\right)^2u.
\end{eqnarray}
We make the following claim~:
\begin{equation}\label{claim1}
\forall\gamma\in\mathbb R,\quad\exists\,C>0,\quad \forall
\tau\in[-1,1],\quad |\xi(\gamma+\tau)-\xi(\gamma)|\leq C|\tau|.
\end{equation} Therefore, thanks to (\ref{Der0-2}) and (\ref{claim1}),
there exist  constants $\tilde C,
\tau_0>0$ such that, for all $\tau\in[-\tau_0,\tau_0]$, we have~:
$$
\left\|\left(H[\gamma+\tau]-\Theta(\gamma)-|\varphi_\gamma(0)|^2\tau\right)u\right\|_{L^2(\mathbb
R_+)}\leq \tilde C\tau^2\|u\|_{L^2(\mathbb R_+)}.
$$
We get now by the spectral theorem the existence of an eigenvalue
$\tilde\Theta(\gamma+\tau)$ of the operator $H[\gamma+\tau]$ that
satisfies the following estimate~:
\begin{equation}\label{zero}
|\tilde\Theta(\gamma+\tau)-\Theta(\gamma)-|\varphi_\gamma(0)|^2\tau|\leq
\tilde C\tau^2,\quad \forall \tau\in[-\tau_0,\tau_0].
\end{equation}
We make now another claim~:
\begin{equation}\label{claim2}
\forall\gamma\in\mathbb R,\quad
\exists\,C_1>0,\quad\forall\tau\in[-1,1],\quad
\left|\mu^{(2)}(\gamma+\tau,\xi(\gamma+\tau))-\mu^{(2)}(\gamma,\xi(\gamma))\right|\leq
C|\tau|, \end{equation} where for $(\eta,\xi)\in\mathbb
R\times\mathbb R$, $\mu^{(2)}(\eta,\xi)$
denotes the second eigenvalue of the operator $H[\eta,\xi]$.\\
Under the above claim, the estimate (\ref{zero}) gives~:
$$\tilde\Theta(\gamma+\tau)=\Theta(\gamma+\tau),\quad \forall \tau\in[-\tau_0,\tau_0].$$
Consequently, we get that $\Theta(\gamma)$ is differentiable and
satisfies formula (\ref{derG}). We make now a final claim~:
\begin{equation}\label{claim3}
\text{The function }\gamma\mapsto|\varphi_\gamma(0)|^2 \text{ is
locally Lipschitz}.
\end{equation}
To achieve the proof of the theorem, we only need to prove
(\ref{claim1}), (\ref{claim2}) and (\ref{claim3}).\\
{\it Proof of (\ref{claim1}).}\\
As we have the formula (\ref{relXiGa}), it is sufficient to prove~:
\begin{equation}\label{Pcl1}
\forall \gamma\in\mathbb R,\quad\exists\,C>0,\quad
\forall\tau\in[-1,1],\quad|\Theta(\gamma+\tau)-\Theta(\gamma)|\leq
C|\tau|.
\end{equation}
The min-max principle gives~:
\begin{equation}\label{Pcl1-1}
\mu^{(1)}(\gamma,\xi)+\tau|\varphi_{\gamma+\tau,\xi}(0)|^2
\leq\mu^{(1)}(\gamma+\tau,\xi)\leq\mu^{(1)}(\gamma,\xi)
+\tau|\varphi_{\gamma,\xi}(0)|^2,\quad \forall\xi\in\mathbb R.
\end{equation}
Thus, given an eigenfunction $\varphi$ of $H[\gamma,\xi]$, we need
to estimate $|\varphi(0)|^2$. Let $u\in D(H[\gamma])$. Using
(\ref{Den-C0H1}) we get~:
\begin{equation}\label{II-36'}
|u(0)|^2\leq 2\|u\|_{L^2(\mathbb R_+)}\|u'\|_{L^2(\mathbb R_+)}.
\end{equation}
We use now (\ref{Th1.2.2-qf}) (with $\epsilon=1/2$) to obtain~:
\begin{equation}\label{II-36''}
\|u'\|_{L^2(\mathbb R_+)}^2\leq 2q[\gamma,\xi](u)
+(\gamma_-)^2\|u\|_{L^2(\mathbb R_+)}^2.\end{equation} Combining
(\ref{II-36'}) and (\ref{II-36''}), we get after an integration by
parts,
\begin{equation}\label{Pcl1-2}
|u(0)|^2\leq 2\left\|H[\gamma,\xi]u\right\|_{L^2(\mathbb
R_+)}\|u\|_{L^2(\mathbb R_+)}+(\gamma_-)^2\|u\|_{L^2(\mathbb
R_+)}^2,\quad \forall u\in D(H[\gamma]).
\end{equation}
Let $M:=\displaystyle\sup_{\tau\in[-1,1]}\xi(\gamma+\tau)$. Let us
show that $M<+\infty$.  Actually, the min-max principle gives~:
$$\mu^{(1)}(\gamma-1,\xi)\leq\mu^{(1)}(\gamma+\tau,\xi)\leq\mu^{(1)}(\gamma+1,\xi),
\quad\forall\tau\in[-1,1],\quad\forall\xi\in\mathbb R.$$ Recalling
(\ref{F-formula}), we obtain
$\Theta(\gamma-1)\leq\displaystyle\sup_{\tau\in[-1,1]}\Theta(\gamma+\tau)\leq\Theta(\gamma+1)$.
Formula (\ref{relXiGa}) now gives $M<+\infty$. Therefore,
(\ref{Pcl1-2}) gives~:
$$|\varphi_{\gamma,\xi}(0)|^2\leq C,\quad
|\varphi_{\gamma+\tau,\xi}(0)|^2\leq C,\quad
\forall\tau\in[-1,1],\quad\forall\xi\in[-M,M],$$ for some constant
$C>0$. Consequently (\ref{Pcl1-1}) yields the estimate~:
$$\mu^{(1)}(\gamma,\xi)-C\tau_-\leq\mu^{(1)}(\gamma+\tau,\xi)\leq\mu^{(1)}(\gamma,\xi)+C
\tau_+,\quad \forall\xi\in[-M,M].$$ Minimizing with respect to
$\xi$, we get (\ref{Pcl1}), thanks to
Theorem~\ref{DaHe}.\\
{\it Proof of (\ref{claim2}).}\\
Let $u\in B^1(\mathbb R_+)$. We shall compare
$q[\gamma+\tau,\xi(\gamma+\tau)](u)$ and $q[\gamma,\xi(\gamma)](u)$.
In fact, we have~:
\begin{eqnarray*}
q[\gamma+\tau,\xi(\gamma+\tau)][u]&=&q[\gamma,\xi(\gamma)](u)
-2(\xi(\gamma+\tau)-\xi(\gamma))\int_0^{+\infty}(t-\xi(\gamma))|u(t)|^2dt\\
&&+(\xi(\gamma+\tau)-\xi(\gamma))^2\int_0^{+\infty}|u(t)|^2dt+\tau|u(0)|^2,
\end{eqnarray*}
where, combining (\ref{II-36'}) and (\ref{II-36''}),
$$|u(0)|^2\leq 2q[\gamma,\xi](u)+(\gamma_-)^2\|u\|^2_{L^2(\mathbb
R_+)}.$$ The Cauchy-Schwarz inequality gives~:
$$2\left|\int_0^{+\infty}(t-\xi(\gamma))|u(t)|^2dt\right|\leq
\int_0^{+\infty}\left|(t-\xi(\gamma))u(t)\right|^2dt+\int_0^{+\infty}|u(t)|^2dt.$$
Using (\ref{claim1}), we get a constant $C>0$ such that, for all
$\tau\in[-1,1]$, we have~:
\begin{eqnarray*}
(1-C\tau_-)q[\gamma,\xi(\gamma)](u)-C\tau_-\|u\|_{L^2(\mathbb
R_+)}^2&\leq&
q[\gamma,\xi(\gamma+\tau)](u)\\
&\leq&(1+C\tau_+)q[\gamma,\xi(\gamma)](u)+C\tau_+\|u\|_{L^2(\mathbb
R_+)}^2. \end{eqnarray*} The min-max principle proves now the
claim.\\
{\it Proof of (\ref{claim3}).}\\
Let $u=\varphi_{\gamma}(t)-e^{-\tau t}\varphi_{\gamma+\tau}(t)$. It
is sufficient to prove that~:
\begin{equation}\label{Lip..}
|u(0)|^2\leq C|\tau|,\quad \forall \tau\in[-\tau_0,\tau_0],
\end{equation}
for constants $C,\tau_0>0$. Using (\ref{Pcl1-2}), we have to
estimate $\|u\|_{L^2(\mathbb R_+)}$ and
$\|H[\gamma,\xi]u\|_{L^2(\mathbb R_+)}$. Let
$f=\left(H[\gamma]-\Theta(\gamma)\right)u$. Then~:
$$f=\left(\Theta(\gamma)-\Theta(\gamma+\tau)\right)u+w,$$ where
$$w(t)=e^{-\tau t}\left(-2\tau\partial_t+2(\xi(\gamma+\tau)-\xi(\gamma))(t-\xi(\gamma))
-(\xi(\gamma+\tau)-\xi(\gamma))^2+\tau^2\right)\varphi_\gamma(t).$$
Therefore, thanks to (\ref{claim1}) and (\ref{Pcl1}), we have~:
\begin{equation}\label{cont-w}
\|f\|_{L^2(\mathbb R_+)}\leq C|\tau|,\quad \forall
\tau\in[-\tau_0,\tau_0]. \end{equation} Noticing that, after an
integration by parts, $\langle
f,\varphi_{\gamma}\rangle_{L^2(\mathbb R_+)}=0$, we write,
$$u=\left(H[\gamma]-\Theta(\gamma)\right)^{-1}f.$$
It is a standard result that the operator norm of
$\left(H[\gamma]-\Theta(\gamma)\right)^{-1}$ is bounded on the
orthogonal space of $\varphi_\gamma$ and is estimated by the inverse
of the gap between the first two eigenvalues of $H[\gamma]$.
Therefore, thanks to (\ref{cont-w}), we get that $\|u\|_{L^2(\mathbb
R_+)}\leq \widetilde C|\tau|$ for some constant $\widetilde C>0$.
Plugging this estimate together with (\ref{cont-w}) in
(\ref{Pcl1-2}), we get (\ref{Lip..}).\hfill$\Box$\\

In the next proposition we have a regularity result with respect to
the two variables $(\gamma,\xi)$.

\begin{prop}\label{1Bonn}
The functions $(\gamma,\xi)\mapsto\mu^{(1)}(\gamma,\xi)$ and
$(\gamma,\xi)\mapsto\varphi_{\gamma,\xi}$ are of class $C^\infty$ in
$\mathbb R^2$. Moreover, we have~:
\begin{equation}\label{1Der-ga-xi}
\partial_\gamma\mu^{(1)}(\gamma,\xi)=|\varphi_{\gamma,\xi}(0)|^2.
\end{equation}
\end{prop}

Using Proposition~\ref{propTay} and Remark~\ref{derThga},
Proposition~\ref{1Bonn} is sufficient for achieving the proof of
Proposition~\ref{reg}.

\noindent{\bf Proof of Proposition~\ref{1Bonn}.} In order to reduce
the problem to a problem of an operator with a fixed domain, we
define the bounded operator $V[\gamma]$ on $L^2(\mathbb R_+)$ by~:
$$ V[\gamma] u=e^{\gamma\chi(t)t}u,\quad \forall u\in L^2(\mathbb R_+).$$
We define then the operator $\tilde H[\gamma,\xi]$ by~:
\begin{eqnarray*}
&&D(\tilde H[\gamma,\xi])=\{u\in B^2(\mathbb R_+); u'(0)=0\},\\
&&\tilde H[\gamma,\xi]=V[-\gamma]H[\gamma,\xi]V[\gamma].
\end{eqnarray*}
Note that the domain of $\tilde H[\gamma,\xi]$ is independent of
$\gamma$ and $\xi$. Note also that  $\tilde H[\gamma,\xi]$ is not
self-adjoint but it has the same spectrum as $H[\gamma,\xi]$. A
fundamental state of $\tilde H[\gamma,\xi]$ is given by~:
$$\tilde\varphi_{\gamma,\xi}(t)=V[-\gamma]\varphi_{\gamma,\xi}(t).$$
We denote by $\varphi_{\gamma,\xi}^*$ the orthogonal projector on
$\varphi_{\gamma,\xi}$. Let us consider a point $(\gamma_0,\xi_0)$.
We define the operator $M_0:D(\tilde H[\gamma,\xi])\times\mathbb
C\longrightarrow L^2(\mathbb R_+)\times \mathbb C$ by~:
$$M_0:=\left(
\begin{array}{cc}
\tilde H[\gamma_0,\xi_0]-\mu_0& \tilde\varphi_0\\
\varphi_0^*&0
\end{array}\right),$$
where $\mu_0=\mu^{(1)}(\gamma_0,\xi_0)$ and
$\varphi_0=\varphi_{\gamma_0,\xi_0}$.\\
The operator $M_0$ is invertible and its inverse $R_0$ is given by~:
$$R_0=\left(\begin{array}{ll}
E_0&E_0^+\\
E_0^-&E_0^{+-}\end{array}\right),$$ where the coefficients of $R_0$
are~:
\begin{eqnarray}
&&E_0=V[-\gamma_0]\tilde R[\gamma_0,\xi_0]V[\gamma_0],\label{1Bon2.10}\\
&&E_0^+=V[-\gamma_0]\varphi_0,\label{1Bon2.11}\\
&&E_0^-=\varphi_0^*V[\gamma_0],\label{1Bon2.12}\\
&&E_0^{+-}=0.\label{1Bon2.13}
\end{eqnarray}
The operator $\tilde R[\gamma_0,\xi_0]$ is the regularized resolvent
which is equal to $0$ on $\mathbb R\cdot\varphi_0$ and to
$\left(H[\gamma_0,\xi_0]-\mu_0\right)^{-1}$ on $\varphi_0^\bot$.\\
Now we define, in a neighborhood of $(\gamma_0,\xi_0,\mu_0)$ the
operator $M(\gamma,\xi,\mu)$ by~:
$$M(\gamma,\xi,\mu)=\left(
\begin{array}{cc}
\tilde H[\gamma,\xi]-\mu&\tilde\varphi_0\\
\varphi^*_0& 0\end{array}\right).$$ The operator $M(\gamma,\xi,\mu)$
is also invertible in a neighborhood of $(\gamma_0,\xi_0,\mu_0)$ and
we denote its inverse by~:
$$R(\gamma,\xi,\mu)=\left(
\begin{array}{ll}
E(\gamma,\xi,\mu)&E^+(\gamma,\xi,\mu)\\
E^-(\gamma,\xi,\mu)&E^{+-}(\gamma,\xi,\mu)\end{array}\right).$$ It
is then standard to prove the following two points (cf. Ref.~\onlinecite{kach} for details)~:\\
$\bullet$ The coefficients of $R(\gamma,\xi,\mu)$ are $C^\infty$ in
a neighborhood of $(\gamma_0,\xi_0,\mu_0)$.\\
$\bullet$ A number $\mu$ is an eigenvalue of $H[\gamma,\xi]$ if and
only if $E^{+-}(\gamma,\xi,\mu)=0$.\\
Moreover, in a neighborhood of $(\gamma_0,\xi_0,\mu_0)$, if $\mu$ is
an eigenvalue of $H[\gamma,\xi]$, then
$V[\gamma]E^+(\gamma,\xi,\mu)$ is a
corresponding eigenfunction.\\
Thus, in a neighborhood of $(\gamma_0,\xi_0)$, the eigenvalues of
the operator $H[\gamma,\xi]$ are given by the solutions of the
equation $E^{+-}(\gamma,\xi,\mu)=0$. By viewing the operator
$M(\gamma,\xi,\mu)$ as a perturbation of $M_0$, we can calculate the
coefficients of $R(\gamma,\xi,\mu)$ and we obtain that~:
$$\partial_\mu E^{+-}(\gamma_0,\xi_0,\mu_0)=1.$$
As the function $E^{+-}(\gamma,\xi,\mu)$ is of class $C^\infty$, we
can apply the implicit function theorem and get the existence of a
number $\eta>0$ and a function $\mu$ of class $C^\infty$ such that~:
\begin{eqnarray*}&&\forall
(\gamma,\xi)\in]\gamma_0-\eta,\gamma_0+\eta[\times]\xi_0-\eta,\xi_0+\eta[,\quad
\forall
\mu\in]\mu_0-\eta,\mu_0+\eta[,\\
&&E^{+-}(\gamma,\xi,\mu)=0\Leftrightarrow
\mu=\mu(\gamma,\xi).\end{eqnarray*} This proves that the functions
$(\gamma,\xi)\mapsto\mu^{(1)}(\gamma,\xi)$ and
$(\gamma,\xi)\mapsto\varphi_{\gamma,\xi}$ are of class
$C^\infty$.\hfill$\Box$\\

\subsection{Asymptotic behavior}
The asymptotic behavior at $\pm\infty$ of the eigenvalue
$\Theta(\gamma)$ with respect to the parameter $\gamma$ is given in
the following proposition.

\begin{prop}\label{propAsy}
There exist  constants $C_0,\gamma_0>0$ such that the eigenvalue
$\Theta(\gamma)$ satisfies~:
\begin{equation}\label{+infty}
1-C_0\gamma\exp(-\gamma^2)\leq\Theta(\gamma)<1,\quad
\forall\gamma\in[\gamma_0,+\infty[,
\end{equation}
and
\begin{equation}\label{-infty}
-\gamma^2\leq\Theta(\gamma)\leq-\gamma^2+ \frac1{4\gamma^2},\quad
\forall \gamma\in]-\infty,0[.
\end{equation}
\end{prop}
\noindent{{\bf Proof.}} We prove the estimate  (\ref{+infty}). Note
that by the min-max principle and Theorem~\ref{DaHe} we get for any
$\gamma>0$~:
\begin{equation}\label{II-25'}
\mu^{(1)}(0,\xi(\gamma))
\leq \Theta(\gamma)<1.
\end{equation}
The following estimate for the
Neumann problem is obtained by Bolley-Helffer~\cite{BoHe} (formula
(A.18))~:
$$\left|\mu^{(1)}(0,\xi)-1\right|\leq C\xi\exp-\xi^2,\quad
\forall\xi\in[A,+\infty[,$$ where $C,A>0$ are constants independent
of $\xi$. Recalling (\ref{relXiGa}), the last estimate gives~:
$$\left|\mu^{(1)}(0,\xi(\gamma))-1\right|\leq
C_0\gamma\exp-\gamma^2,\quad\forall\gamma\in[\gamma_0,+\infty[,$$
where $\gamma_0=\max(\sqrt{A},1)$ and $C_0=2C$. Upon substitution in
(\ref{II-25'}), we arrive at the estimate (\ref{+infty}).\\
The relation (\ref{relXiGa}) gives the lower bound
$\Theta(\gamma)\geq-\gamma^2$. To get the upper bound in
(\ref{-infty}), we use the function $e^{\gamma t}$ (with $\gamma<0$)
as a trial function for the quadratic form defining $H[\gamma,0]$,
this which gives,
$$\frac{q[\gamma,0](e^{\gamma t})}{\|e^{\gamma t}\|_{L^2(\mathbb
R_+)}^2}\leq-\gamma^2+\frac1{4\gamma^2},\quad \forall
\gamma\in]-\infty,0[.$$ Therefore, we get by the min-max principle
that
$\mu^{(1)}(\gamma,0)\leq-\gamma^2+\displaystyle\frac1{4\gamma^2}$.
Recalling (\ref{F-formula}), we get the upper bound in
(\ref{-infty}).
\hfill$\Box$\\

\subsection{Exponential decay of the ground state}
Using Agmon's technique (cf. Ref.~\onlinecite{Agm}), we get the
following decay result for the eigenfunction $\varphi_\gamma$.

\begin{prop}\label{propDecEF}
For each $\epsilon\in]0,1[$ there is a positive constant
$C_\epsilon$ such that, for all $\gamma\in\mathbb R$, we have the
following estimate for the eigenfunction $\varphi_\gamma$~:
\begin{equation}\label{EqDecEF}
\left\|\exp\left(\epsilon\frac{(t-\xi(\gamma))^2}2\right)\varphi_\gamma\right\|_{H^1(\{t\in\mathbb
R_+;(t-\xi(\gamma))\geq C_\epsilon\})}\leq
C_\epsilon(1+\gamma_-+\gamma_-^2),
\end{equation}
where we use the notation $\gamma_-=\max(-\gamma,0)$.
\end{prop}
\noindent{\bf Proof.} Let us consider a function $\Phi\in
H^1(\mathbb R_+)$. Given an integer $N\in\mathbb N$, an integration
by parts gives the following identity~:
\begin{eqnarray}\label{1Agmeqn1-Th}
&&\int_0^N \left[\left|\left(e^\Phi\varphi_\gamma\right)'\right|^2+
\left|(t-\xi(\gamma))e^\Phi\varphi_\gamma\right|^2\right]dt
+\gamma\left|e^{\Phi(0)}\varphi_\gamma(0)\right|^2
-\varphi_\gamma'(N)e^{2\Phi(N)}\varphi_\gamma(N)\nonumber\\
&&=\Theta(\gamma)\left\|e^\Phi\varphi_\gamma\right\|^2_{L^2([0,N])}
+\left\|\Phi'e^\Phi\varphi_\gamma\right\|^2_{L^2([0,N])}.
\end{eqnarray}
Let us recall that the eigenfunction $\varphi_\gamma$ is strictly
positive. It results then from the eigenvalue equation satisfied by
$\varphi_\gamma$~:
$$\varphi_\gamma''(t)=\left((t-\xi(\gamma))^2-\Theta(\gamma)\right)\varphi_\gamma(t)>0,\quad
\forall t\in ]\sqrt{\Theta(\gamma)}+\xi(\gamma),+\infty[.$$
Therefore, the function $\varphi_\gamma'$ is increasing on
$]\sqrt{\Theta(\gamma)}+\xi(\gamma),+\infty[$. On the other hand, as
$\varphi_\gamma\in H^2(\mathbb R_+)$, the Sobolev imbedding theorem
gives $\displaystyle\lim_{t\to+\infty}\varphi_\gamma'(t)=0$. Thus,
combining with the monotonicity of $\varphi_\gamma'$, we get finally
that~:
$$\varphi_\gamma'(t)<0,\quad \forall
t\in]\sqrt{\Theta(\gamma)}+\xi(\gamma),+\infty[.$$ Taking
$N>\sqrt{\Theta(\gamma)}+\xi(\gamma)$ and recalling that
$\Theta(\gamma)<1$, the identity (\ref{1Agmeqn1-Th}) yields the
estimate~:
\begin{eqnarray}\label{1Agmeqn2-Th}
&&\int_0^N \left[\left|\left(e^\Phi\varphi_\gamma\right)'\right|^2+
\left|(t-\xi(\gamma))e^\Phi\varphi_\gamma\right|^2\right]dt
+\gamma\left|e^{\Phi(0)}\varphi_\gamma(0)\right|^2\nonumber\\
&&\leq\left\|e^\Phi\varphi_\gamma\right\|^2_{L^2([0,N])}
+\left\|\Phi'e^\Phi\varphi_\gamma\right\|^2_{L^2([0,N])}.
\end{eqnarray}
To estimate the boundary term in (\ref{1Agmeqn2-Th}), we recall that
(\ref{Pcl1-2}) (with $u=\varphi_\gamma$ and $\xi=\xi(\gamma)$)
gives~:
$$|\varphi_\gamma(0)|^2\leq 2+(\gamma_-)^2.$$
Therefore, the estimate (\ref{1Agmeqn2-Th}) becomes~:
\begin{equation}\label{1Agmeqn3-Th}
\int_0^N \left[\left|\left(e^\Phi\varphi_\gamma\right)'\right|^2+
\left((t-\xi(\gamma))^2-|\Phi'|^2-1\right)\left|e^\Phi\varphi_\gamma\right|^2\right]dt
\leq\gamma_-\sqrt{2+(\gamma_-)^2}e^{2\Phi(0)}.
\end{equation}
Now we take $\Phi$ as~:
$$\Phi(t)=\epsilon\frac{(t-\xi(\gamma))^2}2.$$
We can then rewrite (\ref{1Agmeqn3-Th}) as~:
\begin{equation}\label{1Agmeqn4-Th}
\int_{t\in[0,N],(t-\xi(\gamma))\geq
a_\epsilon}\left[\left|\left(e^\Phi\varphi_\gamma\right)'\right|^2
+\left|e^\Phi\varphi_\gamma\right|^2\right]dt\leq\gamma_-\sqrt{2+(\gamma_-)^2}
e^{\epsilon\xi(\gamma)}+e^{\epsilon a_\epsilon},
\end{equation}
where $a_\epsilon>0$ satisfies~:
$$a_\epsilon^2-\epsilon^2a_\epsilon-1\geq1.$$
Notice that the first term on the right hand side of
(\ref{1Agmeqn4-Th}) is effective only if $\gamma<0$. Coming back to
the regularity of the function $\Theta(\gamma)$, the decay of
$\Theta(\gamma)$ in (\ref{-infty}) and the relation (\ref{relXiGa}),
we get that the function $\xi(\gamma)$ is bounded for $\gamma<0$.
Let us now take~:
$$C_0=\sup_{\gamma<0}\xi(\gamma),\quad
C_\epsilon=\max\left(a_\epsilon,e^{\epsilon C_0},e^{\epsilon
a_\epsilon}\right).$$ The estimate (\ref{1Agmeqn4-Th}) reads now
as~:
$$\int_{t\in[0,N],(t-\xi(\gamma))\geq
C_\epsilon}\left[\left|\left(e^\Phi\varphi_\gamma\right)'\right|^2
+\left|e^\Phi\varphi_\gamma\right|^2\right]dt\leq
C_\epsilon(1+\gamma_-+(\gamma_-)^2).$$ Noticing that the above
estimate is uniform with respect to $N$, we get (\ref{EqDecEF})
upon passing to the limit $N\to+\infty$.\hfill$\Box$\\

Let us now recall that the regularized resolvent $\tilde R[\gamma]$
is the bounded operator defined on $L^2(\mathbb R_+)$ by~:
\begin{equation}\label{1+Resreg}
\tilde R[\gamma]\phi=\left\{
\begin{array}{lll}
0&;& \phi\parallel \varphi_\gamma,\\
\left(H[\gamma]-\Theta(\gamma)\right)^{-1}\phi&;&\phi\bot\varphi_\gamma,
\end{array}\right.
\end{equation}
and extended by linearity. Again, using the Agmon's technique, we
get that this regularized resolvent is uniformly continuous in
suitable weighted spaces.
\begin{prop}\label{propDecReg}
For each $\delta\in]0,1[$ and $\eta_0>0$, there exist positive
constants $C_0,t_0$ such that,
$$\forall\gamma\in[-\eta_0,\eta_0],
\quad\forall u\in L^2\left(\mathbb
R_+;e^{\delta(t-\xi(\gamma))}dt\right),\quad u\bot\varphi_\gamma,$$
we have,
\begin{equation}\label{decRregE}
\left\|e^{\delta(t-\xi(\gamma))}\tilde
R[\gamma]u\right\|_{H^1([t_0,+\infty[)}\leq
C_0\left\|e^{\delta(t-\xi(\gamma))}u\right\|_{L^2(\mathbb R_+)}.
\end{equation}
\end{prop}

\section{\label{Pkach1}Proof of Theorem \ref{kach1}}
In this section we prove Theorem \ref{kach1} by comparing with the
basic model introduced in the preceding section. We introduce a
coordinate system $(s,t)$  near the boundary $\partial\Omega$ where
$t$ measures the distance to $\partial\Omega$  and $s$ measures the
distance in $\partial\Omega$ (cf. Appendix \ref{A}).

\begin{prop}\label{propUB}(Upper bound)\\
Under the hypothesis of Theorem \ref{kach1}, there exist positive
constants $C$ and $h_0$ such that, $\forall h\in]0,h_0]$, we have~:
\begin{equation}\label{Th4.3.1}
\mu^{(1)}(\alpha,\gamma,h)\leq
h\Theta\left(h^{\alpha-1/2}(\gamma_0+Ch^{1/2})\right)+Ch^{3/2}.
\end{equation}
\end{prop}
\noindent{{\bf Proof.}} We start with the easy case when
$\alpha<\frac12$ and $\gamma_0>0$. Notice that in this case, given a
constant $C>0$, formula (\ref{+infty}) gives the existence of
$h_0>0$ such that~:
\begin{equation}\label{+infty'}
\left|\Theta\left(h^{\alpha-1/2}(\gamma_0+Ch^{1/2})\right)-1\right|\leq
\exp(-h^{2\alpha-1}),\quad\forall h\in]0,h_0].
\end{equation}
By comparing with the Dirichlet realization, the min-max principle
gives~:
$$\mu^{(1)}(\alpha,\gamma,h)\leq\lambda^{(1)}(h),$$
where $\lambda^{(1)}(h)$ is the first eigenvalue of the Dirichlet
realization (on $\Omega$) of $-(h\nabla-iA)^2$. Using the following
upper bound for $\lambda^{(1)}(h)$ (cf. Ref.~\onlinecite{HeMo2})~:
$$\lambda^{(1)}(h)\leq h+Ch^{3/2},\quad \forall h\in ]0,1],$$
together with (\ref{+infty'}), we get (\ref{Th4.3.1}).\\
We suppose now that $\gamma_0\leq0$ if $\alpha<\frac12$. Consider a
point $x_0\in\partial\Omega$ such that $\gamma(x_0)=\gamma_0$. We
suppose that $x_0=0$ in the coordinate system $(s,t)$ near the
boundary (cf. Appendix \ref{A}). Using this coordinate system  we
construct a trial function $u_{h,\alpha}$ supported in the rectangle
$K_h=]-h^{1/4},h^{1/4}[\times[0,t_0[$ following the idea of
Helffer-Morame~\cite{HeMo3} and Bernoff-Sternberg~\cite{BeSt}. Since
$x_0$ is a minimum of $\gamma$, Taylor's formula up to the first
order gives the existence of positive constants $C_1,h_0$ such that,
$$\forall h\in]0,h_0],\quad\left|\gamma(s)-\gamma_0\right|\leq C_1h^{1/2}\quad\text{ in }
]-h^{1/2},h^{1/2}[.$$
Thus, given a trial function $u$ supported in $K_h$, we have the
following estimate~:
\begin{equation}\label{3appFQ}
q^{\alpha,\gamma}_{h,A,\Omega}(u)\leq
q^{\alpha,\tilde\gamma_0}_{h,A,\Omega}(u),\quad \forall h\in]0,h_0],
\end{equation}
where $\tilde\gamma_0=\gamma_0+C_1h^{1/2}$. So it is enough to work
with $q^{\alpha,\tilde\gamma_0}_{h,A,\Omega}$.\\
We introduce $\eta=h^{\alpha-1/2}\tilde\gamma_0$ and we choose now
the following trial function~:
\begin{equation}\label{qm322}
u_{h,\alpha}=a^{-1/2}\exp\left(-i\frac{\xi(\eta)
s}{h^{1/2}}\right)v_{h,\alpha},
\end{equation}
where $a(s,t)=1-t\kappa_{\rm r}(s)$ and
\begin{equation}\label{qm323}
v_{h,\alpha}= h^{-3/8} \varphi_\eta\left(h^{-1/2}t\right)
\chi(t)\times f\left(h^{-1/4}s\right).
\end{equation}
The function $\chi$ is a cut-off equal to $1$ in a compact interval
$[0,t_0/2]$ and the function $f\in
C^\infty_0(]-\frac12,\frac12[;\mathbb R)$
is chosen such that $\|f\|_{L^2(\mathbb R)}=1$.\\
Note that the decay of $\varphi_\eta$ in Proposition~\ref{propDecEF}
gives\footnote{Actually we shall need this decay only when
$\alpha<1/2$ and $\gamma_0<0$.}~: For every $\delta>0$ and
$k\in\mathbb N$, there exist positive constants $C_{k,\delta}$ and
$h_0$ such that,
\begin{equation}\label{locBRD}
\int_{\mathbb R_+}t^k|\varphi_\eta(t)|^2dt\leq
C_{k,\delta}h^{-\delta k}, \quad \forall h\in]0,h_0].
\end{equation}
We work with the choice of gauge given in Proposition~\ref{Agd1}.
Using formula (\ref{qfstco}), we can write~:
\begin{eqnarray}\label{ref-add}
&&q^{\alpha,\tilde\gamma_0}_{h,A,\Omega}(u_{h,\alpha}) =
\int_{-|\partial\Omega|/2}^{|\partial\Omega|/2}\int_0^{t_0}
\left[|h\partial_tv_{h,\alpha}|^2
+a^{-2}\left|\left(h^{1/2}\xi(\eta)
-t\left(1-\frac{t}2\kappa_{\rm r}(s)\right)\right)v_{h,\alpha}\right|^2\right]dsdt\nonumber\\
&&\hskip4cm+h^{3/2}\eta\int_{-|\partial\Omega|/2}^{|\partial\Omega|/2}|v_{h,\alpha}(s,0)|^2ds\nonumber\\
&&+h^2\int_{-|\partial\Omega|/2}^{|\partial\Omega|/2}\int_0^{t_0}\left[
|(\partial_ta^{-1/2})v_{h,\alpha}|^2+2a^{-1/2}(\partial_ta^{-1/2})(\partial_tv_{h,\alpha})
v_{h,\alpha}+a^{-2}|\partial_sv_{h,\alpha}|^2\right]a\,dsdt.\nonumber\\
\end{eqnarray}
Recalling the expression of $v_{h,\alpha}$ (cf. (\ref{qm323})), we
can replace the function $\chi$ by $1$ getting an exponentially
small error on the right hand side of (\ref{ref-add}), thanks to the
decay of $\varphi_\eta$ in Proposition~\ref{propDecEF}. After a
change of variables and using the decay of $\varphi_\eta$ in
(\ref{locBRD}), the leading order term on the right hand side of
(\ref{ref-add}) is equal to~:
$$h\left(\int_0^{+\infty}\left[|\varphi'_\eta(t)|^2+|(t-\xi(\eta))\varphi_\eta|^2dt\right]dt
+\eta|\varphi_\eta(0)|^2\right),$$ and the error is of order
$\mathcal O(h^{3/2})$. Therefore, we get constants $C,h_0>0$ such
that~:
$$\left|q^{\alpha,\tilde\gamma_0}_{h,A,\Omega}(u_{h,\alpha})- h\Theta(\eta)\right|\leq Ch^{3/2},\quad
\forall h\in]0,h_0].$$ Using formula (\ref{nostco}) and the decay of
$\varphi_\eta$ (Proposition \ref{propDecEF}), we obtain that the
$L^2$ norm of $u_{h,\alpha}$ is exponentially close to $1$ as
$h\to0$. The application of the min-max principle permits now to
prove (\ref{Th4.3.1}).\hfill$\Box$\\

\begin{rem}\label{eff-gamm}
In the regime $\alpha\in]\frac12,1[$, we have, thanks to
Proposition~\ref{propTay}~:
$$\Theta\left(h^{\alpha-1/2}\gamma_0\right)=\Theta_0+6M_3h^{\alpha-1/2}+\mathcal
O(h^{2\alpha-1}).$$ Substituting the above expansion in the upper
bound (\ref{Th4.3.1}), we get the following upper bound for the
eigenvalue $\mu^{(1)}(\alpha,\gamma,h)$,
$$\mu^{(1)}(\alpha,\gamma,h)\leq h\Theta_0+6M_3\gamma_0h^{\alpha+1/2}+\mathcal O(h^{\inf(3/2,2\alpha)}).$$
We shall prove that this upper bound is actually an asymptotic
expansion of $\mu^{(1)}(\alpha,\gamma,h)$ as $h$ tends $0$ (see
Remark~\ref{alp112}).
\end{rem}

\begin{prop}\label{propLB}(Lower bound)\\
Under the hypothesis of Theorem \ref{kach1}, there exist positive
constants $C,C'$ and $h_0$ such that, $\forall h\in]0,h_0]$, we
have~:
\begin{equation}\label{LB}
\mu^{(1)}(\alpha,\gamma,h)\geq
h\Theta\left(h^{\alpha-1/2}\gamma_0(1+C'h^{1/4})\right)-Ch^{5/4}.
\end{equation}
\end{prop}
\noindent{{\bf Proof.}} We follow the technique of
Ref.~\onlinecite{HeMo3} and we localize by means of a partition of
unity to compare with the model operators in $\mathbb R^2$ and
$\mathbb R\times\mathbb R_+$. Let us explain the heuristic idea. A
partition of unity permits to estimate the quadratic form
$q_{h,A,\Omega}^{\alpha,\gamma}$ locally in small subsets of
$\Omega$. Near the boundary, we obtain after a transformation of
coordinates that the expression of $q_{h,A,\Omega}^{\alpha,\gamma}$
is to leading order asymptotics as that of the half-plane model. In
the interior of $\Omega$, the expression of the quadratic form is
actually like that of the entire-plane model.\\
Let us introduce a partition of unity $(\chi_j)$ of $\mathbb R^2$
that satisfies
$$\sum_j|\chi_j|^2=1,\quad \sum_j|\nabla\chi_j|^2<+\infty,\quad {\rm
supp}\chi_j\subset D(z_j,1),$$ where for $z\in\mathbb R^2$ and
$r>0$, we denote by $D(z,r)$ the disk of center $z$ and
radius $r$.\\
We introduce now the scaled partition of unity~:
$$\chi_j^h(z):=\chi_j(\epsilon_0h^\rho z),\quad \forall z\in\mathbb R^2,$$
where $\epsilon_0$ and $\rho$ are two positive numbers to be chosen
suitably. Note that $(\chi_j^h)$ now satisfies~:
\begin{equation}\label{sum1+}
\sum _j |\chi _j^h|^2=1,
\end{equation}
\begin{equation}\label{lb2+}
\sum _j |\nabla\chi _j^h|^2\leq C\epsilon _0^{-2}h^{-2\rho},
\end{equation}
\begin{equation}\label{support+}
\text{supp }\chi _j^h\subset Q_j^h:=D(z_j^h,\epsilon _0h^\rho),
\end{equation}
where $C$ is a positive constant. We can also suppose that~:
\begin{equation}\label{alternative}
\text{either  supp }\chi _j^h\cap \partial\Omega=\emptyset\quad
\text{or } z_j^h\in \partial\Omega.
\end{equation}
 Note that the alternative in
(\ref{alternative}) permits us to write the sum in  (\ref{sum1+})
under the form~:
$$\sum =\sum _{int}+\sum _{bnd},$$
where the summation over ``$int$'' means that the support of
$\chi_j^h$ do not meet the boundary while that over
``$bnd$'' means the converse.\\
We have now the following decomposition formula~:
\begin{equation}\label{IMS}
q^{\alpha,\gamma} _{h,A}(u)=\sum _jq^{\alpha,\gamma} _{h,A}(\chi
_j^h u) -h^2\sum _j\|\,|\nabla\chi _j^h|\,u\|^2,\quad \forall u\in
H^1(\Omega),
\end{equation}
usually called the IMS formula (cf. Ref.~\onlinecite{CFKS}). We have
now to bound from below each of the terms on the right hand side of
(\ref{IMS}). Note that (\ref{lb2+}) permits to estimate the
contribution of the last term in (\ref{IMS})~:
\begin{equation}\label{clb2}
h^2\sum_j\|\,|\nabla\chi _j^h|\,u\|^2\leq C\epsilon
_0^{-2}h^{2-2\rho}\|u\|^2,\quad \forall u\in H^1(\Omega).
\end{equation}
If $\chi_j^h$ is supported in  $\Omega$, then we have~:
$$q^{\alpha,\gamma}_{h,A}(\chi _j^h u)=\int_{\mathbb
R^2}|(h\nabla-iA)\chi_j^hu|^2dx.$$ Since the lowest eigenvalue of
the Schr\"odinger operator with constant magnetic field in $\mathbb
R^2$ is equal to $h$, we get~:
\begin{equation}\label{c2HeMolbqf}
q^{\alpha,\gamma}_{h,A}(\chi _j^h u)\geq h\int _\Omega |\chi _j^h
u|^2dx,\quad \forall u\in H^1(\Omega).
\end{equation}
We have now to estimate $q^{\alpha,\gamma}_{h,A,\Omega}(\chi_j^hu)$
when $\chi_j^h$ meets the boundary. It is in this case that we see
the effect of the boundary condition. Note that, by writing
$q^{\alpha,\gamma}_{h,A,\Omega}(\chi_j^hu)$ in the boundary
coordinates, thanks to Proposition~\ref{transf}, there exists a
positive constant $C_1$ independent of $h$ and $j$ such that~:
\begin{equation}\label{expbrd}
\int_\Omega|(h\nabla-iA)\chi_j^hu|^2dx\geq
(1-C_1\epsilon_0h^\rho)\int_{\mathbb R\times\mathbb
R_+}|(h\nabla-i\tilde A)\chi_j^hu|^2dsdt,\quad\forall u\in
H^1(\Omega),
\end{equation}
where $\tilde A$ is the vector field associated to $A$ by
(\ref{chmnew}).\\
By a gauge transformation, we get a new magnetic potential $\tilde
A_{new,j}$ satisfying~:
$$\tilde A_{new,j}=\tilde A-\nabla\phi _j^h,$$
$$\tilde A_{new,j}(z_j^h)=0,$$
\begin{equation}\label{linearization}
\left|\tilde A_{new,j}(w)-\tilde A_{lin}^j(w)\right|\leq
C|w|^2,\quad w=(s,t),
\end{equation}
where $A_{lin}^j:=\frac 12 (-t,s)$ is the linear magnetic potential
and  $C>0$ is a constant independent of $h$ and $j$.\\
Given $\theta>0$ and any function $v$ of support in $\mathbb
R\times\mathbb R_+$, we get by the Cauchy-Schwarz inequality,
\begin{eqnarray*}
&&\left|\int_{\mathbb R\times\mathbb R_+}(h\nabla-i\tilde
A_{new,j})v\cdot\overline{(\tilde A_{new,j}-\tilde
A_{lin}^j)v}\,dsdt\right|\\
&&\leq h^{2\theta}\int_{\mathbb R\times\mathbb R_+}|(h\nabla-i\tilde
A_{new,j})v|^2 +h^{-2\theta} \int_{\mathbb R\times\mathbb
R_+}|(\tilde A_{new,j}-\tilde A_{lin}^j)v|^2.
\end{eqnarray*}
Writing $\tilde A_{new,j}=\tilde A_{lin}^j+(\tilde A_{new,j}-\tilde
A_{lin}^j)$ and using (\ref{linearization}), we get  a positive
constant $\tilde C$ independent of $h$ and $j$ such that~:
\begin{equation}\label{eqn-v..}
\int_{\mathbb R\times\mathbb R_+}|(h\nabla-i\tilde
A_{new,j})v|^2dsdt\geq (1-h^{2\theta})\int_{\mathbb R\times\mathbb
R_+}\left|(h\nabla-i\tilde A_{lin}^j) v\right|^2dsdt -\tilde
Ch^{-2\theta}\left\|\,|w|^2\,\chi_j^hu\right\|^2.
\end{equation}
Let us recall that $\chi_j^hu$ is supported in the disk
$D(z_j^h,\epsilon_0h^\rho)$. Upon noticing that
$$\int_{\mathbb
R\times\mathbb R_+}|(h\nabla-i\tilde A)\chi_j^hu|^2dsdt
=\int_{\mathbb R\times\mathbb R_+}\left|(h\nabla-i\tilde
A_{new,j})\exp\left(-i\frac{\phi _j^h}h\right)
\,\chi_j^hu\right|^2dsdt,$$ we get by combining (\ref{eqn-v..})
(with $v=\exp\left(-i\frac{\phi _j^h}h\right) \,\chi_j^hu$) together
with (\ref{expbrd}), a constant $C_2>0$ such that~:
\begin{eqnarray}\label{eqn...es}
\int_{\mathbb R\times\mathbb R_+}|(h\nabla-iA)\chi_j^hu|^2dx&\geq&
(1-C_2\epsilon _0 h^\rho-C_2h^{2\theta})\int_{\mathbb R\times\mathbb
R_+}\left|(h\nabla-i\tilde A_{lin}^j) \exp\left(-i\frac{\phi
_j^h}h\right) \,\chi_j^hu\right|^2dsdt\nonumber\\
&&-C_2\epsilon_0h^{4\rho-2\theta}\int_{\mathbb R\times\mathbb
R_+}|\chi_j^hu|^2dsdt.
\end{eqnarray}
Notice also, (possibly changing $C_2$), we have in
$D(z_j^h,\epsilon_0h^\rho)$,
$$\gamma(x)\geq \gamma(z_j^h)-C_2\epsilon_0h^\rho.$$
Then, by putting,
$$\tilde\gamma_j=
\frac{\gamma(z_j^h)-C_2\epsilon_0h^\rho}{1-C_2h^{2\theta}-C_2\epsilon_0h^\rho},$$
the estimate (\ref{eqn...es}) reads finally~:
\begin{eqnarray}\label{EnLB}
\hskip0.5cm q_{h,A,\Omega}^{\alpha,\gamma}(\chi _j^hu)&\geq&
(1-C_2\epsilon _0 h^\rho-C_2h^{2\theta})q_{h,\tilde
A_{lin}^j,\mathbb R\times\mathbb
R_+}^{\alpha,\tilde\gamma_{j}}\left(\exp\left(-i\frac{\phi
_j^h}h\right)
\chi_j^hu\right)\nonumber\\
&&-C_2\epsilon_0^2h^{4\rho-2\theta}\|\chi _j^hu\|^2.
\end{eqnarray}
Note that this permits to compare with the half-plane model operator
and to get finally the energy estimate (cf. (\ref{HP-alhga}))~:
\begin{equation}\label{EnEst}
q^{\alpha,\gamma}_{h,A,\Omega}(\chi_j^hu)\geq
\left\{(1-C_2\epsilon_0h^\rho-C_2h^{2\theta})
h\Theta(h^{\alpha-1/2}\tilde\gamma_j)-C_2\epsilon_0^2h^{4\rho-2\theta}\right\}
\|\chi_j^hu\|^2_{L^2(\Omega)}.
\end{equation}
 We substitute now the estimates
(\ref{LB}), (\ref{clb2}), (\ref{c2HeMolbqf})  in (\ref{IMS}) and get
finally~:
\begin{eqnarray}\label{weak}
&&q^{\alpha,\gamma}_{h,A,\Omega}(u)\geq h\sum _{int}\int_\Omega
|\chi _j^hu|^2dx +h
\sum_{bnd}\Theta\left(h^{\alpha-1/2}\tilde\gamma_j\right)\int_\Omega
|\chi_j^hu|^2dx\nonumber\\
&&\hskip2cm -C\left(h^{4\rho-2\theta}+\epsilon
_0^{-2}h^{2-2\rho}+h^{1+\rho}+h^{1+2\theta}\right)\|u\|^2,\quad
\forall u\in H^1(\Omega).
\end{eqnarray}
As $\gamma_0$ is the minimum of $\gamma$, we can replace
(\ref{EnLB}) by the estimate~:
\begin{eqnarray}\label{EnLB1}
\hskip0.5cm q_{h,A,\Omega}^{\alpha,\gamma}(\chi _j^hu)&\geq&
(1-C_2\epsilon _0 h^\rho-C_2h^{2\theta})q_{h,\tilde
A_{lin}^j,\mathbb R\times\mathbb
R_+}^{\alpha,\tilde\gamma_0}\left(\exp\left(-i\frac{\phi
_j^h}h\right)
\chi_j^hu\right)\nonumber\\
&&-C_2\epsilon_0^2h^{4\rho-2\theta}\|\chi _j^hu\|^2,
\end{eqnarray}
where $\tilde\gamma_0$ is defined by~:
$$\tilde\gamma_0:=\frac{\gamma_0}{1-C_2h^{2\theta}-C_2\epsilon_0h^\rho}.$$
We then get instead of (\ref{weak})~:
\begin{eqnarray}\label{weak1}
&&q^{\alpha,\gamma}_{h,A,\Omega}(u)\geq h\sum _{int}\int_\Omega
|\chi _j^hu|^2dx +h\Theta\left(h^{\alpha-1/2}\tilde\gamma_0\right)
\sum_{bnd}\int_\Omega
|\chi_j^hu|^2dx\nonumber\\
&&\hskip2cm -C\left(h^{4\rho-2\theta}+\epsilon
_0^{-2}h^{2-2\rho}+h^{1+\rho}+h^{1+2\theta}\right)\|u\|^2,\quad
\forall u\in H^1(\Omega).
\end{eqnarray}
The advantage of (\ref{weak}) is that it gives a lower bound of the
quadratic form $q_{h,A,\Omega}^{\alpha,\gamma}$ in terms of a
potential, see however Section~\ref{Loc}.\\
We choose now $\epsilon_0=1$, and we optimize by taking
$2-2\rho=1+\rho=4\rho-2\theta$ (i.e. $\rho=3/8$ and $\theta=1/8$) in
(\ref{weak1}). We obtain then
(\ref{LB}) by applying the min-max principle.\hfill$\Box$\\

\noindent{\bf Proof of Theorem~\ref{kach1}.} The proof follows in
principle from Propositions~\ref{propUB} and \ref{propLB}. Actually,
in the regime $\alpha<\frac12$, we use further
Proposition~\ref{propAsy}, while in the regime $\alpha\geq\frac12$,
we use the continuity of the function $\Theta(\gamma)$
(Proposition~\ref{reg}).\hfill$\Box$\\


\section{\label{Loc}Localization of the ground state}
We work in this section under the hypotheses of Theorem
\ref{kachLoc}. Due to Theorem~\ref{kach1} we have in this case that
\begin{equation}\label{infSp<1}
\lim_{h\to0}\frac{\mu^{(1)}(\alpha,\gamma,h)}h<1.
\end{equation}
Then this gives, by following the same lines of the proof of
Theorem~6.3 in Ref.~\onlinecite{HeMo3}, the following proposition.
\begin{thm}\label{prop4.1}
Under the hypotheses of Theorem~\ref{kachLoc}, there exist positive
constants $\delta,C,h_0$ such that, for all $h\in]0,h_0]$, a ground
state $u_{\alpha,\gamma,h}$ of the operator
$P_{h,A,\Omega}^{\alpha,\gamma}$ satisfies~:
\begin{equation}\label{eq.4.5+}
\left \|\exp\left( \frac{\delta
d(x,\partial\Omega)}{h^{\beta}}\right)u_{\alpha,\gamma,h}\right\|_{L^2(\Omega)}
\leq C\|u_{\alpha,\gamma,h}\|_{L^2(\Omega)},
\end{equation}
and
\begin{equation}\label{eq.4.7}
\left\|\exp \left(\frac{\delta
d(x,\partial\Omega)}{h^{\beta}}\right) u_{\alpha,\gamma,h}
\right\|_{H^1(\Omega)}\leq
Ch^{-\min(1/2,\beta)}\|u_{\alpha,\gamma,h}\|_{L^2(\Omega)},
\end{equation}
where $\beta=1-\alpha$ if $\gamma_0<0$ and $\alpha<\frac12$, and
$\beta=1/2$ otherwise.
\end{thm}
\noindent{\bf Proof.} Integrating by parts, we get for any Lipschitz
function $\Phi$~:
\begin{eqnarray}\label{AgmIde}
q^{\alpha,\gamma}_{h,A,\Omega}\left(\exp\left(\frac\Phi{h^\beta}\right)
u_{\alpha,\gamma,h}\right)&=&\mu^{(1)}(\alpha,\gamma,h)\left\|\exp\left(\frac\Phi{h^\beta}\right)
u_{\alpha,\gamma,h}\,\right\|^2_{L^2(\Omega)}\\
&&+h^{2-2\beta}\left\||\nabla\Phi|\,\exp\left(\frac\Phi{h^\beta}\right)
\,u_{\alpha,\gamma,h}\right\|^2_{L^2(\Omega)}.\nonumber
\end{eqnarray}
Let $u=\exp\left(\frac\Phi{h^\beta}\right)u_{\alpha,\gamma,h}$.
Using the lower bound for $q^{\alpha,\gamma}_{h,A,\Omega}(u)$ in
(\ref{weak1}) together with  the upper bound for
$\mu^{(1)}(\alpha,\gamma,h)$ in (\ref{Th4.3.1}), we get from
(\ref{AgmIde})~:
\begin{eqnarray*}
&&\hskip-0.5cm\sum_{int}\int_\Omega\left(1-\Theta(h^{\alpha-1/2}(\gamma_0+Ch^{1/2}))
-C\epsilon_0^{-2}h^{1-2\rho}
-Ch^{4\rho-2\theta-1}\right.\\
&&\hskip5cm\left.
-Ch^{\min(\rho,2\theta)}-h^{1-2\beta}|\nabla\Phi|^2\right)\times|\chi_j^hu|^2dx\\
&&\hskip-0.5cm\leq \sum_{bnd}\int_\Omega
\left(\Theta(h^{\alpha-1/2}\tilde\gamma_0)-\Theta(h^{\alpha-1/2}(\gamma_0+Ch^{1/2}))\right.\\
&&\hskip4cm
+\left.h^{\min(4\rho-2\theta-1,1-2\rho)}+h^{1-2\beta}|\nabla\Phi|^2\right)
\times |\chi_j^hu|^2dx.
\end{eqnarray*}
We choose $\rho=\beta$ so that each $\chi_j^h$ is supported in a
disk of radius $\epsilon_0h^\beta$. We choose also $\theta>0$ such
that $4\rho-2\theta-1>0$ and we define the function $\Phi$ by~:
$$\Phi(x)=\delta\max\left({\rm
dist}(x,\partial\Omega);\epsilon_0h^\beta\right),$$ where $\delta$
is a positive constant to be chosen appropriately. Note that
$1-\Theta(h^{\alpha-1/2}\tilde\gamma_0)$ decays in the following
way~:
\begin{eqnarray*}
&&\exists\; C_0,h_0>0\text{ s. t.}, \forall h\in]0,h_0],\\
&&1-\Theta(h^{\alpha-1/2}\tilde\gamma_0)\geq C_0h^{2\alpha-1}\text{
if }\gamma_0<0\text{ and }\alpha<\frac12,\\
&&1-\Theta(h^{\alpha-1/2}\tilde\gamma_0)\geq C_0\text{ otherwise.}
\end{eqnarray*}
Thus we can choose $\epsilon_0$ and $\delta$ small enough, so that
we get finally the following decay~:
$$\sum_{int}\int_\Omega\left|\chi_j^h\exp\frac{\Phi}{h^\beta}\,u_{\alpha,\gamma,h}\right|^2dx
\leq C\int_\Omega|u_{\alpha,\gamma,h}|^2dx.$$ This actually permits
to conclude (\ref{eq.4.5+}) and, thanks to (\ref{AgmIde}),
\begin{equation}\label{eq.4.5'}
q^{\alpha,\gamma}_{h,A,\Omega}\left(\exp\frac{\delta
d(x,\partial\Omega)}{h^{\beta}}u_{\alpha,\gamma,h}\right)\leq
Ch^{\min(2-2\beta,1)}\left\|\exp\frac{\delta
d(x,\partial\Omega)}{h^{\beta}}u_{\alpha,\gamma,h}\right\|^2_{L^2(\Omega)}.\end{equation}
For a function $u\in H^1(\Omega)$, let $\tilde u(s,t)$ be defined by
means of boundary coordinates $(s,t)$ and equal to the restriction
of $u$ in $\Omega_{t_0}$ (cf. Appendix~\ref{A}). Notice that~:
$$|\tilde u(s,0)|^2=
-2\int_0^\infty \left\{\partial_t\left(\chi(t)\tilde
u(s,t)\right)\right\}\chi(t) \tilde u(s,t)dt,$$ where $\chi$ is the
same cut-off introduced in (\ref{qm323}). Integrating the above
identity with respect to the variable $s$ then applying a
Cauchy-Schwarz inequality, we get after a change of variables the
following interpolation inequality~:
$$\|u\|^2_{L^2(\partial\Omega)}\leq
C\|u\|_{L^2(\Omega)}\times\|u\|_{H^1(\Omega)},$$ where $C$ is a
positive constant depending only on $\Omega$.\\
Applying again a Cauchy-Schwarz inequality, the preceding estimate
gives~:
$$\|(h\nabla-iA)u\|^2_{L^2(\Omega)}\leq
2q^{\alpha,\gamma}_{h,A,\Omega}(u)+Ch\|u\|^2_{L^2(\Omega)},\quad
\forall u\in H^1(\Omega).$$ In particular, for
$u=\exp\left(\frac{\delta
d(x,\partial\Omega)}{h^\beta}\right)u_{\alpha,\gamma,h}$, we get
(\ref{eq.4.7}), thanks to (\ref{eq.4.5+}) and
(\ref{eq.4.5'}).\hfill$\Box$\\

We study now the decay near the boundary. Let us consider a number
$\beta>0$ and a Lipschitz function  $\Phi_0$ defined in
$\overline{\Omega}$. The function $\Phi_0$ and the number $\beta$
will be chosen later in an appropriate manner. Choosing
$\rho=\frac38$, $\theta=\frac18$ and $\epsilon_0$ large enough, the
energy estimate (\ref{weak}) together with the upper bound
(\ref{Th4.3.1}) give the existence of a positive constant $C$ such
that~:
\begin{eqnarray*}
&&0\geq h\sum _{int}\int_\Omega
\left(1-\Theta\left(h^{\alpha-1/2}\tilde\gamma_0\right)
-Ch^{1/4}-h^{1-2\beta}|\nabla\Phi_0|^2\right)
\left|\exp \left(\frac{\Phi _0}{h^{\beta}}\right)\chi_j^hu_{\alpha,\gamma,h}\right|^2dx\\
&&\hskip0.7cm+h\sum _{bnd}\int_\Omega
\left[\left(\Theta\left(h^{\alpha-1/2}\tilde\gamma(x)\right)
-\Theta\left(h^{\alpha-1/2}\tilde\gamma_0\right)\right)-Ch^{1/4}-h^{1-2\beta}|\nabla\Phi
_0|^2\right]\\
&&\hskip2.8cm\times \left |\exp
\left(\frac{\Phi_0}{h^{\beta}}\right)\chi_j^hu_{\alpha,\gamma,h}\right|^2dx,\nonumber
\end{eqnarray*}
where $\tilde\gamma_0=\gamma_0+Ch^{1/2}$. The function $\gamma$ is
extended to a small boundary sheath by means of boundary coordinates
in the following way~:
$$\gamma(x)=\gamma(s(x)),\quad \forall x\in\Omega_{t_0}.$$
In the case $\alpha<\frac12$ and  $\gamma_0=0$, thanks to
Proposition \ref{propAsy}, the difference between
$\Theta\left(h^{\alpha-1/2}\tilde\gamma(x)\right)$ and
$\Theta\left(h^{\alpha-1/2}\tilde\gamma_0\right)$ decays in the
following way~:
$$\forall \varepsilon>0,\,\exists C_\varepsilon>0,\,\,\forall x\in(\gamma-\gamma_0)^{-1}([\epsilon,+\infty[),\,\,
\Theta\left(h^{\alpha-1/2}\tilde \gamma(x)\right)
-\Theta\left(h^{\alpha-1/2}\tilde\gamma_0\right)>C_\varepsilon.$$ In
the case $\alpha<\frac12$ and $\gamma_0<0$, we have a stronger
decay~:
\begin{eqnarray*}
&&\forall \varepsilon>0,\,\exists C_\varepsilon>0,\,\,\forall
x\in(\gamma-\gamma_0)^{-1}([\epsilon,+\infty[),\,\,\\
&&\Theta\left(h^{\alpha-1/2}\tilde
\gamma(x)\right)-\Theta\left(h^{\alpha-1/2}\tilde\gamma_0\right)>C_\varepsilon
h^{1-2\alpha}.
\end{eqnarray*}
 So by taking $\Phi_0$ in the form~:
$$\Phi_0(x)=\delta\chi({\rm dist}(x,\partial\Omega)){\rm dist}\left(x,\{x\in\partial\Omega;\gamma(x)=\gamma_0\}\right),$$
with $\delta$ an appropriate positive constant and $\chi$ is the
same as in (\ref{qm323}), we get for each $\epsilon>0$ the following
decay near the boundary~:
\begin{equation}\label{eq.4.19}
\int _{{\rm dist}(x,\partial\Omega)<t_0}\left| \exp\frac{\Phi
_0}{h^{\beta}}\,u_{\alpha,\gamma,h}\right|^2dx\leq C_\epsilon
\exp\frac\epsilon{h^{\beta}}\|u_{\alpha,\gamma,h}\|^2,\quad \forall
h\in]0,h_\epsilon],
\end{equation}
with $\beta=1-\alpha$ if $\gamma_0<0$ and $\alpha<\frac12$, and
$\beta=\frac12$ otherwise.
This gives finally the decay in Theorem~\ref{kachLoc}.\\
For the critical case $\alpha=\frac12$ and $\gamma_0$ arbitrary, we
define the function $\Phi_0$ by~:
$$\Phi_0(x)=
\delta\;\chi({\rm dist}(x,\partial\Omega)) \;{\rm dist}_{{\rm
agm}}\left(x,\{x\in\partial\Omega;\gamma(x)=\gamma_0\}\right),$$
where ${\rm dist}_{{\rm agm}}$ is the Agmon distance associated to
the metric $\left(\Theta(\gamma(x))-\Theta(\gamma_0\right))_+$.
We obtain then  a similar decay result to (\ref{eq.4.19}).\\
In the case when $\alpha>\frac12$, we need a finer energy estimate
than (\ref{weak}), see however Remark~\ref{alp112}.

\section{\label{curv}Two-term asymptotics}

In this section we suppose in addition to the hypotheses of
Theorem~\ref{kach1} that $\alpha\geq\frac12$. We give two-term
asymptotic expansions for the ground state energy showing the
influence of the scalar curvature and we finish the proofs of the
remaining theorems announced in the introduction.

\subsection{Upper bound}
We construct a trial function defined by means of boundary
coordinates $(s,t)$ near a point $z_0\in\partial\Omega$. We suppose
that $z_0=0$ in the coordinate system $(s,t)$ and we denote by
$\kappa_0=\kappa_{\rm r}(0)$, $a_0=1-t\kappa_0$ and
$\eta(z_0)=h^{\alpha-1/2}\gamma(z_0)$. We then define the trial
function~:
\begin{equation}\label{cuv-qm}
u_h=\exp\left(-i\frac{\xi(\eta(z_0))s}{h^{1/2}}\right)v_h(s,t),
\end{equation}
with
\begin{equation}\label{curv-qmD}
v_h(s,t)=h^{-5/16}a_0^{-1/2}(t)\varphi_{\eta(z_0)}\left(h^{-1/2}t\right)\chi(t)\cdot
f\left(h^{-1/8}s\right),
\end{equation}
and where the functions $\chi$ and $f$ are as in (\ref{qm323}).\\
We continue now to work in the spirit of Ref.~\onlinecite{HeMo3}. We
work with the gauge given in Proposition~\ref{Agd1}. An explicit
calculation, thanks to the decay of $\varphi_{\eta(z_0)}$
(Proposition \ref{propDecEF}), gives the following lemma.

\begin{lem}\label{HeMo10.26}
With the above notations, for each $\alpha\in[\frac12,1]$ and
$\gamma\in C^\infty(\partial\Omega;\mathbb R)$, there exist positive
constants $C,h_0$ such that, $\forall h\in]0,h_0]$, we have the
following estimate~:
\begin{equation}\label{EHeMo10.25}
\left| q^{\alpha,\gamma(z_0)}_{h,A,\Omega}(u_h)- \int_{\mathbb R_+}
H^h\left(U^h\varphi_{\eta(z_0)}\right)\times
\left(U^h\varphi_{\eta(z_0)}\right)dt\right|\leq Ch^{13/8},
\end{equation}
where the operators $H^h$ and $U^h$ are defined respectively by~:
$$H^h=a_0^{-2}
\left(t\left(1-t\frac{\kappa_0}2\right)-h^{1/2}\xi(\eta(z_0))\right)^2
-h^2a_0^{-1}\partial_t(a_0\partial_t),$$
$$(U^hg)(t)=h^{-1/4}g(h^{-1/2}t),\quad \forall g\in L^2(\mathbb
R_+).$$
\end{lem}
\noindent{\bf Proof.} Note that in the support of $u_h$ we
have\footnote{Actually, if $z_0$ is a point of maximum of
$\kappa_{\rm r}$, the remainder is better and of order $\mathcal
O(h^{3/4})$ for the first term.}
$$a=a_0+\mathcal O(h^{5/8}),\quad \tilde A_1=-t\left(1-\frac t2\kappa_0\right)
+\mathcal O(h^{9/8}).$$ Then, thanks to formula (\ref{qfstco}) (also
cf. (\ref{ref-add})) and the decay of $\varphi_{\eta(z_0)}$
(Proposition~\ref{propDecEF}), we get modulo $\mathcal
O(h^{13/8})$~:
\begin{eqnarray}\label{det-ref..}
q^{\alpha,\gamma(z_0)}_{h,A,\Omega}(u_h)&=& \int_{\mathbb
R\times\mathbb
R_+}a_0\left\{|h\partial_tv_h|^2+a_0^{-2}\left|\left(t\left(1-\frac
t2\kappa_0\right)-h^{1/2}\xi(\eta(z_0))\right)v_h\right|^2\right\}dsdt\nonumber\\
&&+h^{3/2}\eta(z_0)\int_{\mathbb R}|v_h(s,0)|^2.
\end{eqnarray}
Integrating with respect to $s$, the right hand side above is equal
to~:
\begin{eqnarray*}
&&\hskip-0.5cm h^{-1/2}\int_{\mathbb R}a_0\left\{h^2
\left|\partial_t\left(a_0^{-1/2}\varphi_{\eta(z_0)}(h^{-1/2}t)\chi(t)\right)\right|^2\right.\\
&&\left.+a_0^{-3}\left|\left(t\left(1-\frac t2\kappa_0\right)
-h^{1/2}\xi(\eta(z_0))\right)\varphi_{\eta(z_0)}(h^{-1/2}t)\chi(t)\right|^2\right\}dsdt+
h\eta(z_0)|\varphi_{\eta(z_0)}(0)|^2. \end{eqnarray*} We can replace
the function $\chi$ in the above expression by $1$ getting an
exponentially small error, thanks to Proposition~\ref{propDecEF}.
Thus, modulo a small exponential error, we rewrite the above
expression as~:
\begin{eqnarray*}
&&\int_{\mathbb R}\left\{h^2
a_0\left|\partial_t\left(U^h\varphi_{\eta(z_0)}\right)\right|^2
+a_0^{-2}\left|\left(t\left(1-\frac t2\kappa_0\right)
-h^{1/2}\xi(\eta(z_0))\right)\left(U^h\varphi_{\eta(z_0)}\right)\right|^2\right\}dsdt\\
&&+h^{3/2}\eta(z_0)\left|\left(U^h\varphi_{\eta(z_0)}\right)(0)\right|^2.\end{eqnarray*}
Notice that we have the boundary condition
$(U^h\varphi_{\eta(z_0)})'(0)=h^{-1/2}\eta(z_0)(U^h\varphi_{\eta(z_0)})(0)$.
Therefore, integrating by parts, the above expression is equal to
$\int_{\mathbb R_+} H^h\left(U^h\varphi_{\eta(z_0)}\right)\times
\left(U^h\varphi_{\eta(z_0)}\right)dt$. Upon substituting in
(\ref{EHeMo10.25}), this
finishes the proof of the lemma.\hfill$\Box$\\

\noindent Similar computations give also the following lemma.
\begin{lem}\label{LHeMo10.27}
Under the hypotheses of Lemma~\ref{HeMo10.26}, there exist positive
constants $C,h_0$ such that, $\forall h\in]0,h_0]$, we have~:
\begin{equation}\label{EHeMo10.27}
\left\|(H^h-H_0^h-H_1^h)U^h\varphi_{\eta(z_0)}\right\|_{L^2(\mathbb
R_+)}\leq Ch^2,\end{equation} where the operators $H^h_0$ and
$H^h_1$ are defined respectively by~:
$$\begin{array}{l}
H_0^h=-h^2\partial_t^2+\left(t-h^{1/2}\xi(\eta(z_0))\right)^2,\\
\\H_1^h=2t\kappa_0\left(t-h^{1/2}\xi(\eta(z_0))\right)^2
-\kappa_0t^2\left(t-h^{1/2}\xi(\eta(z_0))\right)+h^2\kappa_0\partial_t.
\end{array}$$
\end{lem}

Let us denote by (cf. (\ref{M-3-12ga}) and (\ref{M3}))~:
$$M_3\left(\frac12,\gamma(z_0)\right)=M_3(\gamma(z_0)),\quad
M_3(\alpha,\gamma(z_0))=M_3\text{ for }\alpha>\frac12.$$ The next
lemma permits to conclude an upper bound for the eigenvalue
$\mu^{(1)}(\alpha,\gamma,h)$.
\begin{lem}\label{UB-alp-gam-gen}
Under the above notations, there exist positive constants $C,h_0$
such that, when $h\in]0,h_0]$, we have the following estimate~:
$$\left|q^{\alpha,\gamma}_{h,A,\Omega}(u_h)
-\{\Theta(\eta(z_0))-2M_3(\alpha,\gamma(z_0))\kappa_0h^{3/2}\}\|u_h\|^2_{L^2(\Omega)}\right|
\leq Ch^{\epsilon_\alpha},$$ where
$\epsilon_\alpha=\inf(13/8,2\alpha+\frac12)$ for $\alpha>\frac12$
and $\epsilon_{1/2}=13/8$.
\end{lem}
\noindent{\bf Proof.} Notice that in the support of $u_h$ we have
$\gamma(z)=\gamma(z_0)+\mathcal O(h^{1/8})$. Then this gives~:
$$q^{\alpha,\gamma}_{h,A,\Omega}(u_h)-q^{\alpha,\gamma(z_0)}_{h,A,\Omega}(u_h)=
\mathcal O(h^{9/8+\alpha}).$$ In view of Lemmas~\ref{HeMo10.26} and
\ref{LHeMo10.27}, we get the following estimate~:
\begin{equation}\label{UB-en}
\left|q^{\alpha,\gamma}_{h,A,\Omega}(u_h)-\int_{\mathbb
R_+}\left(H^h_1+H^h_0\right)\left(U^h\varphi_{\eta(z_0)}\right)\times
U^h\varphi_{\eta(z_0)}dt\right|\leq Ch^{13/8}.\end{equation} We note
also that we have the following relations~:
\begin{eqnarray}\label{HeMo10.27+}
&& (U^h)^\star
H_0^hU^h=h\left\{-\partial_t^2+\left(t-\xi(\eta(z_0))\right)^2\right\},\\
&&(U^h)^\star
H_1^hU^h=\kappa_0h^{3/2}H_1,\label{HeMo10.27++}\end{eqnarray} where
the operator $H_1$ is defined by~:
$$
H_1=\left(t-\xi(\eta(z_0))\right)^3-
\xi(\eta(z_0))^2\left(t-\xi(\eta(z_0))\right)+\partial_t.
$$
By defining $K_3(\alpha,h) :=\int_{\mathbb
R_+}H_1\varphi_{\eta(z_0)}\cdot\varphi_{\eta(z_0)}dt$, the estimate
(\ref{UB-en}) reads as~:
\begin{equation}\label{UB-en1}
\left|q^{\alpha,\gamma}_{h,A,\Omega}(u_h)
-\left\{h\Theta(\eta(z_0))+K_3(\alpha,h)\kappa_0h^{3/2}\right\}\right|\leq
Ch^{13/8}.
\end{equation}
Now, for $\alpha=\frac12$, we get by using (\ref{Mn-gam}) that
$K_3(\frac12,h)=-2M_3(\frac12,\gamma(z_0))$. For $\alpha>\frac12$,
thanks to Propositions~\ref{reg} and \ref{propTay}, we get that
$$K_3(\alpha,h)=-2M_3+\mathcal O(h^{2\alpha-1}).$$
Finally, the decay of $\varphi_{\eta(z_0)}$ in
Proposition~\ref{propDecEF} gives that $\|u_h\|_{L^2(\Omega)}$ is
exponentially close to $1$. This achieves the proof of the
lemma.\hfill$\Box$

The min-max principle gives now, thanks to
Lemma~\ref{UB-alp-gam-gen}, an upper bound for
$\mu^{(1)}(\alpha,\gamma,h)$. Under the hypothesis of
Theorem~\ref{kach-thm-alp=1}, we take $z_0$ such that
$$(\kappa_{\rm r}-3\gamma)(z_0)=(\kappa_{\rm r}-3\gamma)_{\rm max}$$
and we use the expansion (cf. (\ref{der0}))~:
$$\Theta(\eta(z_0))=\Theta_0+6M_3\gamma(z_0)h^{1/2}+\mathcal O(h).$$
Therefore, (\ref{UB-en1}) gives the following upper bound~:
\begin{equation}\label{upp-bnd-curv}
\mu^{(1)}(1,\gamma,h)\leq h\Theta_0-2M_3(\kappa_{\rm
r}-3\gamma)_{\rm max}h^{3/2}+\mathcal O(h^{13/8}).
\end{equation}
Under the hypothesis of Theorem \ref{kachLocGC}, we choose $z_0$
such that $\kappa_{\rm r}(z_0)=(\kappa_{\rm r})_{\rm max}$.

\subsection{Lower bound}
As in the proof of Proposition~\ref{propLB}, we consider a standard
scaled partition of unity\footnote{We take a partition of unity
associated to squares instead of discs.}
$\left(\chi_{j,h^{1/6}}\right)_{j\in\mathbb Z^2}$ of $\mathbb R^2$
that satisfies~:
\begin{equation}\label{sum}
\sum _{j\in J}|\chi _{j,h^{1/6}}(z)|^2=1,\quad \sum _{j\in J}|\nabla
\chi _{j,h^{1/6}}(z)|^2\leq Ch^{-1/3},
\end{equation}
\begin{equation}\label{support1}
\text{supp }\chi _{j,h^{1/6}}\subset jh^{1/6}+[-h^{1/6},h^{1/6}]^2.
\end{equation}
We define the following set of indices~:
$$
J^1_{\tau(h)}:=\{j\in \mathbb Z^2; {\rm supp}\chi _{j,h^{1/6}}\cap
\Omega\not=\phi, {\rm dist}(\text{supp } \chi
_{j,h^{1/6}},\partial\Omega)\leq \tau(h) \},
$$
where the number $\tau(h)$ is defined by~:
\begin{equation}\label{eq.5.26}
\tau(h)=h^\delta,\quad \text{with}\quad \frac 16\leq\delta\leq\frac
12,
\end{equation}
and the number $\delta$ will be chosen in a suitable manner.\\
We consider also another scaled partition of unity in $\mathbb R$~:
\begin{equation}\label{CHeMo9.22}
\psi_{0,\tau(h)}^2(t)+\psi_{1,\tau(h)}^2(t)=1,\quad|\psi'_{j,\tau(h)}(t)|\leq\frac{C}{\tau(h)}
,\quad j=0,1,\end{equation}
\begin{equation}\label{CHeMo9.23}
{\rm
supp}\,\psi_{0,\tau(h)}\subset[\frac{\tau(h)}{20},+\infty[,\quad
{\rm
supp}\,\psi_{1,\tau(h)}\subset]-\infty,\frac{\tau(h)}{10}].\end{equation}
Note that, for each $j\in J^1_{\tau(h)}$, the function
$\psi_{1,\tau(h)}(t)\chi_{j,h^{1/6}}(s,t)$ could be interpreted, by
means of boundary coordinates, as a function in $\overline{\Omega}$.
Moreover, each $\psi_{1,\tau(h)}(t)\chi_{j,h^{1/6}}(s,t)$ is
supported in a rectangle
$$K(j,h)=]-h^{1/6}+s_j,s_j+h^{1/6}[\times
[0,h^{\delta}[$$ near $\partial\Omega$. The role of $\delta$ is then
to control the size of the width of each rectangle $K(j,h)$. Due to
the exponential decay of a ground state away from the boundary
(Theorem \ref{prop4.1}), we get the following lemma.
\begin{lem}\label{lemKach}
Suppose that $\alpha>\frac12$. With the above notations, a
$L^2$-normalized ground state $u_{\alpha,\gamma,h}$ of the operator
$P_{h,A,\Omega}^{\alpha,\gamma}$ satisfies~:
\begin{equation}\label{eq.5.29}
\left| \sum _{j\in J^1_{\tau(h)}} q^{\alpha,\gamma}_{h,A,\Omega}
\left(\chi
_{j,h^{1/6}}\psi_{1,\tau(h)}u_{\gamma,h}\right)-\mu^{(1)}(\alpha,\gamma,h)
\right|\leq Ch^{5/3}.
\end{equation}
\end{lem}
\noindent The proof of (\ref{eq.5.29}) follows the same lines of
that in
Ref.~\onlinecite{HeMo3} (Formulas (10.4), (10.5) and (10.6)).\\
\noindent For each $j\in J^1_{\tau(h)}$, we define a unique point
$z_j\in\partial\Omega$ by the relation $s(z_j)=s_j$. We denote then
by $\kappa_j=\kappa_{\rm r}(z_j)$, $a_j(t)=1-\kappa_jt$,
$A^j(t)=-t\left(1-\frac{t}2\kappa_j\right)$, and $\gamma_j=\gamma(z_j)$.\\
We consider now the $k$-family of one dimensional differential
operators~:
\begin{equation}\label{eq.5.31}
H_{h,j,k}=-h^2a_j^{-1}\partial_t(a_j\partial_t)+(1+2\kappa_jt)(hk-A^j)^2,
\end{equation}
where $k$ is a real parameter. We denote by
$H^{\alpha,\gamma,D}_{h,j,k}$ the self-adjoint realization on
$L^2\left(]0,h^\delta[; a_j(t)dt\right)$ of $H_{h,j,k}$ whose domain
is given by~:
\begin{equation}\label{eq.5.31bis}
D(H^{\gamma,D}_{h,j,k})=\{ v\in H^2(]0,h^\delta[);
v'(0)=h^\alpha\tilde\gamma_j v(0), \, v(h^\delta)=0\}.
\end{equation}
The parameter $\tilde\gamma_j$ is defined by~:
$$\tilde\gamma_j=\gamma_j+\varepsilon(h),$$
where $\varepsilon(h)=0$ if the function $\gamma$ is constant; if
$\gamma$ is not constant, then there are constants $C,h_0>0$ such
that~:
$$|\varepsilon(h)|\leq Ch^{1/6},\quad \forall h\in]0,h_0].$$
We now  introduce~:
\begin{equation}\label{eq.5.32}
\mu ^j_1(\alpha,\gamma,h):= \inf _{k\in \mathbb R} \inf {\rm Sp}
(H^{\gamma,D} _{h,j,k}).
\end{equation}
We have now the following lemma.
\begin{lem}\label{lemm5.6}
For each $\alpha\in[\frac12,+\infty[$, we have under the above
notations~:
\begin{equation}\label{eq.5.33}
\mu^{(1)}(\alpha,\gamma,h) \geq \left(\inf _{j\in J^1_{\tau(h)}} \mu
_1^j(\alpha,\gamma,h)\right)+\mathcal O(h^{5/3}).
\end{equation}
\end{lem}
Again the proof follows the same lines of Ref.~\onlinecite{HeMo3}
(Section 11), but let us explain briefly the main steps. We express
each term
$q^{\alpha,\gamma}_{h,A,\Omega}(\psi_{1,\tau(h)}\chi_{j,h^{1/6}}u_{\alpha,\gamma,h})$
in boundary coordinates. We work with the local choice of gauge
given in Proposition~\ref{Agd1}. We expand now all terms by Taylor's
formula  near $(s_j,0)$. After controlling the remainder terms,
thanks to the exponential decay of the ground states away from the
boundary, we apply a partial Fourier transformation in the
tangential variable
$s$ and we get finally the result of the lemma.\\

We have now to find, uniformly over $k\in\mathbb R$, a lower bound
for the first eigenvalue $\mu_1^j(k;\alpha,\gamma,h)$ of the
operator $H^{\alpha,\gamma,D}_{h,j,k}$. Putting $\beta=\kappa_j$,
$\xi=-h^{1/2}k$ and $\eta=\tilde\gamma_j$, we get by a scaling
argument~:
$$\mu_1^j(k;\alpha,\gamma,h)=h\mu_1(H_{h,\beta,\xi}^{\alpha,\eta,D}),$$
where $\mu_1(H_{h,\beta,\xi}^{\alpha,\eta,D})$ is the first
eigenvalue of the one dimensional operator~:
\begin{eqnarray}\label{eq.5.45}
&&H_{h,\beta,\xi}^{\alpha,\eta,D}=-\partial_t^2+(t-\xi)^2+\beta h^{1/2}(1-\beta h^{1/2}t)^{-1}\partial_t
\nonumber\\
&&\hskip0.7cm+ 2\beta h^{1/2}t\left(t-\xi-\beta
h^{1/2}\frac{t^2}2\right)^2 -\beta h^{1/2}t^2(t-\xi)+\beta^2h
\frac{t^4}4,
\end{eqnarray}
whose domain is defined by~:
\begin{eqnarray*}
D(H^{\alpha,\eta,D}_{h,\beta,\xi})=\{u\in H^2(]0,h^{\delta-1/2}[);
u'(0)=h^{\alpha-1/2}\eta\, u(0),\,\, u(h^{\delta-1/2})=0\}.
\end{eqnarray*}
We have then to find (when $\eta,\beta\in]-M,M[$ and $M$ a given
positive constant), uniformly with respect to $\xi\in\mathbb R$, a
lower bound for the eigenvalue
$\mu_1(H_{h,\beta,\xi}^{\alpha,\eta,D})$. The min-max principle
gives the following preliminary localization of the spectrum of the
operator $H_{h,\beta,\xi}^{\alpha,\eta,D}$~:
\begin{lem}\label{lemm5.10}
For each  $M>0$ and $\alpha\in[\frac12,+\infty[$, there exist
positive constants $C,h_0$ such that,
$$\forall\eta,\beta\in]-M,M[,\quad\forall\xi\in\mathbb R,\quad \forall
h\in]0,h_0],$$ we have,
\begin{equation}\label{eq.5.53}
\left|\mu _j(H^{\alpha,\eta,D}_{h,\beta,\xi})-\mu
_j(H^{\alpha,\eta,D}_{0,\xi})\right| \leq
Ch^{2\delta-1/2}\left(1+\mu _j(H^{\alpha,\eta,D}_{0,\xi})\right),
\end{equation}
where, for an operator $T$ having a compact resolvent, $\mu _j(T)$
denotes the increasing sequence of eigenvalues of $T$.
\end{lem}

\begin{rem}\label{remRed}
Note that the min-max principle gives now that
$$\mu _j(H^{\alpha,\eta,D}_{0,\xi})\geq\mu^{(j)}(h^{\alpha-1/2}\eta,\xi),$$
where, for $\tilde\eta\in\mathbb R$, $\mu^{(j)}(\tilde\eta,\xi)$ is
the increasing sequence of eigenvalues of the operator $H[\eta,\xi]$
introduced in (\ref{xiOp}).
\end{rem}

The following lemma deals with the case when $\xi$ is not localized
very close to $\xi(h^{\alpha-1/2}\eta)$.
\begin{lem}\label{lemm5.9}
Suppose that $\delta\in ]1/4,1/2[$. For each $\alpha\geq\frac12$,
there exists $\rho\in]0,\delta-\frac14]$, and for each $M>0$, there
exist positive constants $\zeta,h_0>0$ such that,
$$\forall\eta,\beta\in]-M,M[,\quad \forall\xi \text{ such that }
|\xi-\xi(h^{\alpha-1/2}\eta)|\geq \zeta h^{\rho},\quad\forall
h\in]0,h_0],$$ we have,
\begin{equation}\label{eq.5.48}
\mu_1(H^{\alpha,\eta,D}_{h,\beta,\xi})\geq
\Theta(h^{\alpha-1/2}\eta)+ h^{2\rho}.
\end{equation}
\end{lem}
\noindent{\bf Proof.} It is sufficient to obtain (\ref{eq.5.48}) for
$\mu^{(1)}(h^{\alpha-1/2}\eta,\xi)$, thanks to Lemma~\ref{lemm5.10}
and Remark~\ref{remRed}. We start with the case when
$\alpha=\frac12$ and $\eta\in]-M,M[$. Writing Taylors formula up to
the second order for the function $\xi\mapsto\mu^{(1)}(\eta,\xi)$,
we get positive constants $\theta,C_1$ such that
when~$|\xi-\xi(\eta)|\leq\theta$, we have~:
$$\mu^{(1)}(\eta,\xi)\geq \Theta(\eta)+C_1|\xi-\xi(\eta)|^2.$$
Then by taking $\zeta$ such that $C_1\zeta>\zeta_0$, where
$\zeta_0>1$ is a constant to be chosen appropriately, we get when
$\zeta h^\rho\leq|\xi-\xi(\eta)|\leq\theta$,
$$\mu^{(1)}(\eta,\xi)\geq\Theta(\eta)+\zeta_0h^{2\rho},$$
where $\rho$ is also a positive constant to be chosen later. When
$|\xi-\xi(\eta)|>\theta$, we get a positive constant
$\epsilon_\theta$ such that~:
$$\mu^{(1)}(\eta,\xi)\geq \Theta(\eta)+\epsilon_\theta.$$
Then by choosing $h_0$ such that $\zeta_0h_0^\rho<\epsilon_\theta$,
we get for $|\xi-\xi(\eta)|\geq \zeta h^\rho$ and $h\in]0,h_0]$~:
\begin{equation}\label{eta=con}
\mu^{(1)}(\eta,\xi)\geq \Theta(\eta)+\zeta_0h^{2\rho}.
\end{equation}
We treat now the case when $\alpha>1/2$. Note that the min-max
principle gives uniformly for all $\xi\in\mathbb R$ and
$\eta\in]-M,M[$,
$$\mu^{(1)}(h^{\alpha-1/2}\eta,\xi)\geq
(1-C\eta_-h^{\alpha-1/2})\mu^{(1)}(0,\xi).$$ Then using
(\ref{eta=con}) for $\eta=0$ and
$\rho=\inf(\delta-\frac14,\alpha-\frac12)$, we can choose $\zeta_0$
large enough so that we have for $|\xi-\xi_0|\geq\zeta h^\rho$~:
$$\mu^{(1)}(h^{\alpha-1/2}\eta,\xi)\geq\Theta(h^{\alpha-1/2}\eta)+\frac{\zeta_0}2h^{\rho}.$$
To finish the proof, we replace $\xi_0$ by $\xi(h^{\alpha-1/2}\eta)$
getting an error of order $\mathcal
O(h^{\alpha-1/2})$.\hfill$\Box$\\

Now we deal with the case when $|\xi-\xi(h^{\alpha-1/2}\eta)|<\zeta
h^{\rho}$. Let $\tilde\eta=h^{\alpha-1/2}\eta$. We look for a formal
solution $\left(\mu,f_{h,\beta,\xi}^{\alpha,\eta}\right)$ of the
spectral problem
\begin{equation}\label{6.2.59}
H_{h,\beta,\xi}^{\alpha,\eta}f_{h,\beta,\xi}^{\alpha,\eta} =\mu
f_{h,\beta,\xi}^{\alpha,\eta},\quad
\left(f_{h,\beta,\xi}^{\alpha,\eta}\right)'(0)=h^{\alpha-1/2}f_{h,\beta,\xi}^{\alpha,\eta}(0),
\end{equation}
in the form~:
\begin{eqnarray}\label{VPFML}
&&\hskip-0.5cm\mu=d_0+d_1\left(\xi-\xi(\tilde\eta)\right)
+d_2\left(\xi-\xi(\tilde\eta)\right)^2+d_3 h^{1/2},\\
&&\hskip-0.5cm
f_{h,\beta,\xi}^{\alpha,\eta}=u_0+\left(\xi-\xi(\tilde\eta)\right)u_1
+\left(\xi-\xi(\tilde\eta)\right)^2u_2+h^{1/2}u_3,\label{FPFML}
\end{eqnarray}
where the coefficients $d_0,d_1,d_2,d_3$ and the functions
$u_0,u_1,u_2,u_3$ are to be determined. We expand the operator
$H_{h,\beta,\xi}^{\alpha,\eta,D}$ in powers of
$(\xi-\xi(\tilde\eta))$ and then we identify the coefficients of the
terms of orders $(\xi-\xi(\tilde\eta))^j$ ($j=0,1,2$) and $h^{1/2}$.
We then obtain for the coefficients~:
\begin{equation}\label{d3u3}
\left\{\begin{array}{l}
d_0=\Theta(\tilde\eta),\quad u_0=\varphi_{\tilde\eta}\\
\\
d_1=0,\quad u_1=2\tilde R[\tilde\eta]\left\{(t-\xi(\tilde\eta))\varphi_{\tilde\eta}\right\}\\
\\
d_2=:d_2(\alpha,\eta)=1-2\int_{\mathbb R_+}(t-\xi(\tilde\eta))\varphi_{\tilde\eta}u_1dt,\\
\\
u_2=\tilde R[\tilde\eta]\left\{
4\left(t-\xi(\tilde\eta)\right)\tilde
R[\tilde\eta]\left[\left(t-\xi(\tilde\eta)\right)
\varphi_{\tilde\eta}\right] -d_2\right\}\\
\\
d_3=:d_3(\alpha,\eta)=\beta\int_{\mathbb
R_+}\varphi_{\tilde\eta}\left\{\partial_t
+(t-\xi(\tilde\eta))^3\right\}\varphi_{\tilde\eta}dt,\\
\\
u_3= -\tilde R[\tilde\eta]
\left[\beta\left(\partial_t+\left(t-\xi(\tilde\eta)\right)^3
-\xi(\tilde\eta)^2\left(t-\xi(\tilde\eta)\right)\right)
-d_3\right]u_0.
\end{array}\right.\end{equation}
Using the function
$\chi(\frac{t}{h^{\delta-1/2}})f_{h,\beta,\xi}^{\alpha,\eta}$ (where
$\chi$ is the same as in (\ref{qm323})) as a quasi-mode, we get by
the spectral theorem, thanks to the decay results in Propositions
\ref{propDecEF} and \ref{propDecReg} and to the localization of the
spectrum in Lemma~\ref{lemm5.10}, the following lemma.
\begin{lem}\label{appmuG}
Suppose that $\delta\in]\frac14,\frac12[$. For each $M>0$ and
$\alpha\in[\frac12,1]$, there exist positive constants $C>0,h_0$
such that,
$$\forall \eta,\beta\in]-M,M[,\quad
\forall\xi\text{ such that }|\xi-\xi(\tilde\eta)|\leq\zeta
h^{\rho},\quad\forall h\in]0,h_0],$$ we have,
\begin{eqnarray}\label{EqappmuG}
&&
\left|\mu_1(H_{h,\beta,\xi}^{\alpha,\eta,D})-\left\{\Theta(\tilde\eta)
+d_2(\alpha,\eta)(\xi-\xi(\tilde\eta))^2+
d_3(\alpha,\eta)h^{1/2}\right\}
\right|\\
&&\leq
C\left[h^{1/2}|\xi-\xi(\tilde\eta)|+h^{\delta+1/2}\right].\nonumber
\end{eqnarray}
where $d_2(\alpha,\eta)$ and $d_3(\alpha,\eta)$ are defined by
(\ref{d3u3}) respectively.
\end{lem}
\noindent Hence we have obtained by this analysis a lower bound for
the first eigenvalue $\mu^{(1)}(\alpha,\gamma,h)$. We complete the
picture by showing that the term $d_2(\alpha,\eta)$ is positive.
\begin{lem}
For each $\alpha\in[\frac12,+\infty[$ and $M>0$, there exists a
positive constant $h_0$ such that~:
$$d_2(\alpha,\eta)>0,\quad \forall h\in]0,h_0],\quad\forall \eta\in]-M,M[.$$
\end{lem}
\noindent{\bf Proof.} It is actually sufficient to prove the
conclusion of the lemma when $\alpha=\frac12$. If $\alpha>\frac12$,
we replace $d_2(\alpha,\eta)$ by its approximation up to the first
order, thanks to Proposition~\ref{reg}, and we obtain that~:
$$d_2(\alpha,\eta)=d_2(\frac12,0)+\mathcal O(h^{\alpha-1/2}),$$
which gives the lemma.\\
For the particular case $\alpha=\frac12$, we show that~:
$$d_2(\frac12,\eta)=\frac12\left(\partial_\xi^2\mu^{(1)}(\eta,\cdot)\right)(\xi(\eta))$$
which is strictly positive.\hfill$\Box$\\

We are now able to conclude the asymptotics given in
Theorems~\ref{kach-thm-alp=1} and \ref{kachLocGC}. First we choose
$\delta=\frac5{12}$. When $\alpha>\frac12$ we replace
$\Theta(\tilde\eta)$ and $d_3$ by their approximations  up to the
second and first orders respectively, thanks to
Propositions~\ref{propTay} and \ref{reg}. For $\alpha=\frac12$
we get by (\ref{M3-gam}) that  $d_3$ is indeed equal to $-2M_3(\frac12,\gamma)$.\\

\begin{rem}\label{alp112}
When $\alpha\in]\frac12,1[$ and when the function $\gamma$ is not
constant, we get from the above analysis that  the upper bound in
Remark~\ref{eff-gamm} is indeed an asymptotic expansion, and we
achieve therefore the proof of Theorem~\ref{alpha12-1}.\\
We get also that the quadratic form $q_{h,A,\Omega}^{\alpha,\gamma}$
can be bounded from below by means of a potential $W$~:
$$q^{\alpha,\gamma}_{h,A,\Omega}(u)\geq \int_\Omega W(x)|u(x)|^2dx,\quad \forall u\in H^1(\Omega),$$
where $W$ is defined for some positive constant $C_0$ by~:
$$W(x)=\left\{\begin{array}{clll}h&;&if&{\rm dist}(x,\partial\Omega)>h^{1/6}\\
h\Theta_0+6M_3\gamma(x)h^{\alpha+1/2}-C_0h^{\inf(3/2,2\alpha)}&;&if&{\rm
dist}(x,\partial\Omega)<h^{1/6}.\end{array}\right.$$ Then, as in
Section~\ref{Loc}, we get by Agmon's technique that a ground state
decays exponentially away from the boundary points where $\gamma$ is
minimum and hence we have completed the proof of
Theorem~\ref{kachLoc}.
\end{rem}

\begin{rem}\label{LocGamCon}
Note also that the above analysis permits, under the hypotheses of
Theorems~\ref{kach-thm-alp=1} and \ref{kachLocGC}, to bound the
quadratic form $q^{\alpha,\gamma}_{h,A,\Omega}$ from below using a
potential $W$ defined either by means of the function $\kappa_{\rm
r}-3\gamma$ (when $\alpha=1$) or by the scalar curvature
$\kappa_{\rm r}$ (when $\gamma$ is constant). Then, by using Agmon's
technique, we finish the proofs of Theorems~\ref{kach-thm-alp=1} and
\ref{kachLocGC}.
\end{rem}

\section{Conclusion}
The systematic analysis in the spirit of Ref.~\onlinecite{HeMo3} has
allowed us to understand the role of the boundary condition imposed
by De\,Gennes. We have extended in Theorems~\ref{kach-thm-alp=1} and
\ref{kachLocGC} the expansion announced by Pan \cite{Pa03} in the
particular case when $\alpha=1$ and $\gamma$ is a positive constant.
However, there is a specific difficulty when $\gamma$ is negative.
We have not been able to obtain the localization of the ground state
when $\alpha<1/2$ and $\gamma_0>0$. This is strongly related to the
question of the localization of the ground state of the Dirichlet
realization of the Schr\"odinger operator with constant magnetic
field which is open. Finally, in the spirit of
Refs.~\onlinecite{FoHe1, HePa, LuPa1}, we hope to apply this
analysis to the onset of superconductivity and to complete the
analysis of Ref.~\onlinecite{LuPa96}
(cf. Ref.~\onlinecite{kach}).\\

\section*{Acknowledgements.} I am deeply grateful to Professor B.
Helffer for the constant attention to this work,  his help, advices
and comments. I would like also to thank S. Fournais for his
attentive reading and suggestions, and the referee for his valuable
comments which improved the presentation of the paper. I acknowledge
the ESI at Vienna where I found good conditions to prepare a part of
this work and the ESF which had supported the visit under the SPECT
program. This work has been carried out by the financial support of
the {\it Agence universitaire de la francophonie} (AUF).

\appendix
\section{\label{A}Coordinates near the boundary}
We recall in this appendix well-known coordinates that straightens a
portion of the boundary $\partial\Omega$. Let
$s\in]-\frac{|\partial\Omega|}2,\frac{|\partial\Omega|}2]\mapsto
M(s)\in\partial\Omega$ be a regular parametrization of
$\partial\Omega$. For each $x\in\Omega$ and $\epsilon>0$ we denote
by~:
$$t(x)={\rm dist}(x,\partial\Omega)\text{ and }\Omega_\epsilon=\{x\in\overline{\Omega};{\rm dist}(x,\partial\Omega)<\epsilon\}.$$
Then there exist a positive constant $t_0>0$ depending on $\Omega$
such that, for each $x\in\Omega_{t_0}$, we can define the
coordinates $(s(x),t(x))$ by~:
$$t(x)=|x-M(s(x))|,$$
and such that the transformation~:
$$\psi:\Omega_{t_0}\ni x\mapsto(s(x),t(x))\in\mathbb S_{|\partial\Omega|/2\pi}^1\times[0,t_0[$$
is a diffeomorphisim. The Jacobian of this coordinate transformation
is given by~:
\begin{equation}\label{Jac}
a(s,t)={\rm det}(D\psi)=1-t\kappa_{\rm r}(s).
\end{equation}
To a vector field $A=(A_1,A_2)\in C^\infty(\overline{\Omega};\mathbb
R^2)$, we associate the vector field $\tilde A=(\tilde A_1,\tilde
A_2)\in C^\infty(\mathbb S_{|\partial\Omega|/2\pi}^1\times[0,t_0[)$
by the following relation~:
\begin{equation}\label{chmnew}
\tilde A_1ds+\tilde A_2dt=A_1dx_1+A_2dx_2.
\end{equation}
We get then the following change of variable formulas.
\begin{proposition}\label{transf}
Let $u\in H^1(\Omega)$ be supported in $\Omega _{t _0}$. Then we
have~:
\begin{equation}\label{qfstco}
\int_{\Omega_{t_0}}\left|(h\nabla-iA)u\right|^2dx=\int _{\mathbb
S^1_{|\partial\Omega|/2\pi}\times [0,t_0[}\left[ |(h\partial
_t-i\tilde A_2)v|^2+a^{-2}|(h\partial _s-i\tilde
A_1)v|^2\right]a\,dsdt.
\end{equation}
and
\begin{equation}\label{nostco}
\int_{\Omega_{t_0}} |u(x)|^2\,dx=\int_{\mathbb
S^1_{|\partial\Omega|/2\pi}\times[0,\epsilon_0[} |v(s,t)|^2a\,dsdt,
\end{equation}
where  $v(s,t)=u(\psi^{-1}(s,t))$.
\end{proposition}

We have also the relation~:
$$\left(\partial_{x_1}A_2-\partial_{x_2}A_1\right)dx_1\wedge dx_2=
\left(\partial_s\tilde A_2-\partial_t\tilde A_1\right)a^{-1}ds\wedge
dt,$$ which gives,
$${\rm curl}\, \tilde A=\left(1-t\kappa_{\rm r}(s)\right){\rm curl}\,A.$$
We give in the next proposition a standard choice of gauge.
\begin{prop}\label{Agd1}
Consider a vector field $A=(A_1,A_2)\in
C^\infty(\overline{\Omega};\mathbb R^2)$ such that ${\rm curl }A=1$.
For each point $x_0\in\partial\Omega$, there exist a neighborhood
$\mathcal V_{x_0}\subset\Omega_{t_0}$ of $x_0$  and a smooth
real-valued function $\phi_{x_0}$ such the vector field
$A_{new}:=A-\nabla\phi_{x_0}$ satisfies~:
\begin{equation}\label{HeMo}
\tilde A_{new}^1=-t\left(1-\frac t2\kappa_{\rm r}(s)\right)\text{
and } \tilde A^2_{new}=0\text{ in }\mathcal V_{x_0},
\end{equation}
with $\tilde A_{new}=(\tilde A_{new}^1,\tilde A_{new}^2)$.
\end{prop}

\end{document}